\documentclass[%
 reprint,
 amsmath,amssymb,longbibliography
 aps,
 prl,
]{revtex4-1}

\usepackage{graphicx}
\usepackage{amsmath}
\usepackage{rotating}
\usepackage{color}
\usepackage{dcolumn}
\usepackage{bm}
\usepackage{tikz}
\usetikzlibrary{positioning,calc}
\usetikzlibrary{trees}
\usetikzlibrary{decorations.pathmorphing}
\usetikzlibrary{decorations.markings}
\usetikzlibrary{arrows,positioning} 
\tikzset{
    photon/.style={decorate, decoration={snake}, draw=black},
    electron/.style={draw=black, postaction={decorate},
        decoration={markings,mark=at position .55 with {\arrow[draw=black,thick]{>}}}},
    gluon/.style={decorate, decoration={snake},draw=black}, 
    >=stealth',
    punkt/.style={
           rectangle,
           rounded corners,
           draw=black, very thick,
           text width=6.5em,
           minimum height=2em,
           text centered},
    pil/.style={
           ->,
           thick,
           shorten <=2pt,
           shorten >=2pt,}
}

\newcommand{\sruo}{Sr$_2$RuO$_4$}
\newcommand{\lsoc}{\lambda_{soc}}
\newcommand{\io}{\tilde \mu}

\newcommand{\down}{\downarrow}
\newcommand{\up}{\uparrow}
\newcommand{\spin}{\sigma}
\newcommand{\ospin}{\overline{\sigma}}

\newcommand{\pv}{{\bf p}}
\newcommand{\kv}{{\bf k}}
\newcommand{\qv}{{\bf q}}
\newcommand{\Qv}{{\bf Q}}
\begin{document}

\title{Knight Shift and Leading Superconducting Instability From Spin Fluctuations in \sruo}
\author{A. T. R\o mer,$^{1,2}$ D. D. Scherer,$^1$ I. M. Eremin,$^3$ P. J. Hirschfeld,$^4$ B. M. Andersen$^1$}
\affiliation{%
$^1$ Niels Bohr Institute, University of Copenhagen, Vibenhuset, Lyngbyvej 2, DK-2100 Copenhagen,
Denmark\\
$^2$Institut Laue-Langevin, 
71 avenue des Martyrs CS 20156, 38042 Grenoble Cedex 9, France \\
$^3$ Institut f\"ur Theoretische Physik III, Ruhr-Universit\"at Bochum, D-44801 Bochum, Germany \\
$^4$Department of Physics, University of Florida, Gainesville, Florida 32611, USA
}%

\date{\today}

\begin{abstract}
Recent nuclear magnetic resonance studies [A. Pustogow {\it et al.}, arXiv:1904.00047] have challenged the prevalent chiral triplet pairing scenario proposed for Sr$_2$RuO$_4$.   To provide guidance from microscopic theory as to which other pair states might be compatible with the new data, we perform   a detailed theoretical study of spin-fluctuation mediated pairing  for this compound. We map out the phase diagram as a function of spin-orbit coupling, interaction parameters, and band-structure properties over physically reasonable ranges, comparing when possible with photoemission and inelastic neutron scattering data information.  We find that even-parity pseudospin singlet solutions dominate large regions of the phase diagram, but in certain regimes spin-orbit coupling favors a  near-nodal odd-parity triplet superconducting state, which is either helical or chiral depending on the proximity of the $\gamma$ band to the van Hove points. A surprising near-degeneracy of the nodal $s^\prime$- and  $d_{x^2-y^2}$-wave solutions leads to the possibility of a near-nodal time-reversal symmetry broken  $s^\prime+id_{x^2-y^2}$ pair state.  Predictions for the temperature dependence of the Knight shift for fields in and out of plane are presented for all states.
\end{abstract}

\maketitle

Superconductivity in Sr$_2$RuO$_4$ remains largely a mystery despite the relative simplicity of the material as compared to the high-T$_c$ cuprates and almost twenty five years of intense research efforts\cite{Mackenzie2017}. Until recently, the dominant opinion was that Sr$_2$RuO$_4$ represents a  unique example of a chiral triplet
superconducting state, supported by  the presumed proximity of layered Sr$_2$RuO$_4$ to ferromagnetism\cite{Sigrist1999}, observed in the perovskite ``parent" material SrRuO$_3$, as well as temperature independent  Knight shift data across T$_c$, measured on both  Ru\cite{Ishida97,Ishida15} and O\cite{Ishida98,Manago16} nuclei.  It was soon discovered, however, that the leading magnetic instability in Sr$_2$RuO$_4$ occurs in an antiferromagnetic, and not ferromagnetic channel\cite{Mazin1997,Sidis99,Braden02}, although later weak low $q$-fluctuations were also observed\cite{Braden04,Steffens19}. In this case, the usual spin-fluctuation exchange pairing mechanism\cite{Berk1966}
would be expected to lead to even parity spin-singlet solutions rather than odd parity spin-triplet states. The situation is further complicated by the multi-orbital nature of the electronic states\cite{Oguchi1995,Mazin1997}, as well as sizeable spin-orbit coupling\cite{Ng2000,Annett2006,Zabolotnyy13,Veenstra2014}, resulting in significant magnetic anisotropy of the spin fluctuations in this material\cite{Eremin2002,Braden04,Cobo16}, which 
complicate theoretical analysis.
Furthermore, as the main belief was that Sr$_2$RuO$_4$ supported a spin-triplet superconducting state, most theories focused on such solutions. For a review of earlier works see e.g. Ref. \onlinecite{Mackenzie2017}, and also more recent works, Refs.\cite{Raghu2010,Wang2013,Scaffidi2014,Zhang18,Wang19}.

Very recently, the Knight shift in an in-plane magnetic field was re-measured by a different group and found to drop below $T_c$,  severely challenging the prevalent chiral triplet pair state proposed for Sr$_2$RuO$_4$\cite{Pustogow19}.
Previous results were interpreted as a result of heating of the sample during the application of high amplitude radio-frequency pulses~\cite{Pustogow19}.  Although it is prudent to wait for confirmation of this result, it appears as though the problem of superconductivity in Sr$_2$RuO$_4$ is ripe for reexamination.

In this Letter we present a detailed theoretical study of spin-fluctuation mediated pairing relevant for Sr$_2$RuO$_4$ using 
a realistic spin-orbit coupling (SOC), which correctly reproduces the magnetic anisotropy found in this system, and sizeable Hund's coupling strength\cite{Kim2018}. 
\begin{figure*}[t]
 \centering
   	\includegraphics[angle=0,width=\linewidth]{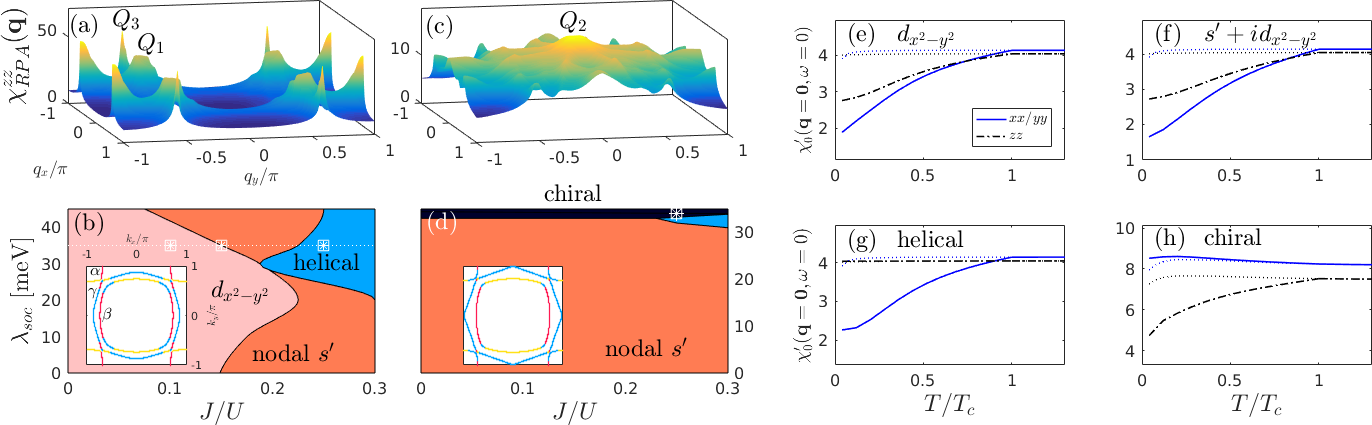}
\caption{Longitudinal spin susceptibility $\chi_{RPA}^{zz,\prime}({\bf q},\omega=0)$ at $\lsoc=35$ meV and leading superconducting instability as a function of SOC amplitude $\lsoc$ and Hund's coupling $J$ for $\mu_{xy}=109$ meV (a,b) and $\mu_{xy}=134$ meV (c,d). The Fermi surface with the $\alpha,\beta$ and $\gamma$ band is shown for each case by insets with dominating orbital content displayed by colors: $xy$-orbital is blue, $xz$ is red and $yz$ is yellow.
In (b,d) white symbols display the positions for which the Knight shifts shown in (e-h) were obtained. The Knight shift is given by $\chi_0^\prime(\qv=0,\omega=0)$, and we set $k_BT_c=0.5$ meV and the maximum amplitude of the gap is $\Delta_{\rm max}=1$ meV. The solid blue line is the Knight shift for in-plane fields ($xx/yy$-component), while the black dashed-dotted line displays the Knight shift for out-of-plane fields ($zz$-component). The dotted lines display the normal state Knight shift $xx/yy$-component (blue) and $zz$-component (black).
}
\label{fig:LGE_Cobo}
 \end{figure*}
In particular, we investigate the leading superconducting instabilities in a framework where SOC is included both in the electronic structure and the pairing interaction. Throughout, we relate our results to neutron scattering data, and additionally discuss the Knight shift and the existence of nodal gaps in the DOS. Finally, we address the role of electron interactions beyond the random phase approximation (RPA) on the preferred Cooper pairing. 

Atomic spin-orbit coupling, parametrized here by $H_{SOC}=\lsoc \bf{L}\cdot\bf{S}$, does not break time-reversal symmetry and due to Kramer's theorem all energies thus remain  doubly degenerate. Degenerate eigenvectors are labelled by pseudo-spin $\sigma=+/-$ and the relation to electronic annihilation/creation operators $c_{\mu,s}(\kv)/c^\dagger_{\mu,s}(\kv)$ of orbital character $\mu$ and spin $s$ is given by
$\Psi(\kv,+)=[c_{xz\up}(\kv),c_{yz,\up}(\kv),c_{xy,\down}(\kv)]$, and 
$\Psi(\kv,-)=[c_{xz,\down}(\kv),c_{yz,\down}(\kv),c_{xy,\up}(\kv)]$.
In this basis the non-interacting Hamiltonian can be written in block-diagonal form $\hat{H}=\sum_\sigma \Psi^\dagger(\kv,\sigma) (H_0+H_{SOC}) \Psi(\kv,\sigma)$ with the matrices $H_0$ and $H_{SOC}$ given by
\begin{eqnarray}
 H_0&=&\left( \begin{array}{ccc}
  \xi_{xz}(\kv) & g(\kv) &0 \\
  g(\kv) & \xi_{yz}(\kv) & 0 \\
  0 & 0 &   \xi_{xy}(\kv)
 \end{array}\right),
\end{eqnarray}
\begin{eqnarray}
 H_{SOC}&=&\frac{1}{2}\left( \begin{array}{ccc}
  0 & -i\spin\lsoc & i\lsoc \\
  i\spin\lsoc & 0 & -\spin\lsoc \\
  -i\lsoc & -\spin\lsoc &  0\\
 \end{array}\right),
 \label{eq:H0Hsoc}
\end{eqnarray}
with $\spin=+(-)$ for pseudo-spin up (down) block. 
The electronic dispersions are given by
$\xi_{xz}(\kv)=-2t_1\cos k_x -2t_2\cos k_y -\mu$,
$\xi_{yz}(\kv)=-2t_2\cos k_x -2t_1\cos k_y -\mu$, and
$\xi_{xy}(\kv)=-2t_3(\cos k_x +\cos k_y) 
-4t_4\cos k_x \cos k_y-2t_5(\cos 2k_x +\cos 2k_y) -\mu_{xy}  
$.
As in Ref.~\onlinecite{Cobo16} we parametrize the band by $\{t_1,t_2,t_3,t_4,t_5\}=\{88,9,80,40,5\}$ meV 
with $g(\kv)=0$ and the chemical potential of the $xz,yz$ orbitals $\mu=109$ meV. Below, $\mu_{xy}$ is allowed to vary slightly from $\mu$ to map out the effect of a different crystal field, motivated by a sensitivity of the superconducting instability to  the proximity of the $xy$ orbital Fermi surface states to the van Hove saddle points. We restrict ourselves to a purely two-dimensional electronic model, given the strong electronic anisotropy of \sruo. Although the third dimension may play a role, the main physics is expected to occur in the RuO$_2$ planes.

We derive the effective electron-electron interaction in the Cooper channel from the multi-orbital Hubbard Hamiltonian which includes intra- and interorbital Coulomb interactions and Hund's coupling terms.
Summation of all ladder and bubble diagrams 
gives the effective interaction expressed in terms of the bare interaction parameters $U,U',J,J'$ and the RPA spin susceptibilities, for more details see Supplementary Material (SM)~\cite{suppl}. This procedure results in the interaction Hamiltonian
\begin{equation}
 \hat{H}_{int}=\frac{1}{2}\!\sum_{ \kv,\kv' \{\tilde \mu\}}\!\!\Big[V(\kv,\kv')\Big]^{\io_1 , \io_2 }_{\io_3,\io_4 }  c_{\kv \io_1 }^\dagger  c_{-\kv \io_3 }^\dagger c_{-\kv' \io_2 } c_{\kv' \io_4 },
 \label{eq:Heff}
\end{equation}
with the pairing interaction given by
\begin{eqnarray}
\Big[V(\kv,\kv')\Big]^{\io_1 , \io_2 }_{\io_3,\io_4 } &=&\Big[U\Big]^{\io_1 , \io_2 }_{\io_3,\io_4 }+\Big[U\frac{1}{1-\chi_0U}\chi_0U\Big]^{\io_1 \io_2}_{\io_3 \io_4}(\kv+\kv') \nonumber \\
&& -\Big[U\frac{1}{1-\chi_0U}\chi_0U  \Big]^{\io_1\io_4}_{\io_3 \io_2}(\kv-\kv') .
\label{eq:Veff}
\end{eqnarray}
The label $\io \:= (\mu,s)$ is a joint index for orbital and electronic spin and $\chi_0=[\chi_0]^{\tilde \mu_1,\tilde \mu_2}_{\io_3,\io_4}(\qv,i\omega_n=0)$ denotes the real part of the static
generalized multi-orbital spin susceptibility in the presence of SOC.
The interaction Hamiltonian as stated in Eq.~(\ref{eq:Heff}) is projected to band and pseudo-spin space to obtain the final form:
\begin{eqnarray}
 \hat{H}_{int} \!=\!\!\!\!\!\!\!\sum_{n,n', \kv,\kv'}\!
 \sum_{l,l'} 
\overline\Psi_{l}(n,\kv)
 ~\!\frac{1}{2}\Gamma_{l,l'}(n,\kv;n',\kv')~\Psi_{l'}(n',\kv').\nonumber \\
\end{eqnarray}
Here $n,n'$ are band indices, and the pseudo-spin information is carried by the $l,l'$ indices 
with the fermion bilinear operator, $\Psi_l(n,\kv)$, defined in SM~\cite{suppl}.
\begin{figure*}[t]
 \centering
    	\includegraphics[angle=0,width=.95\linewidth]{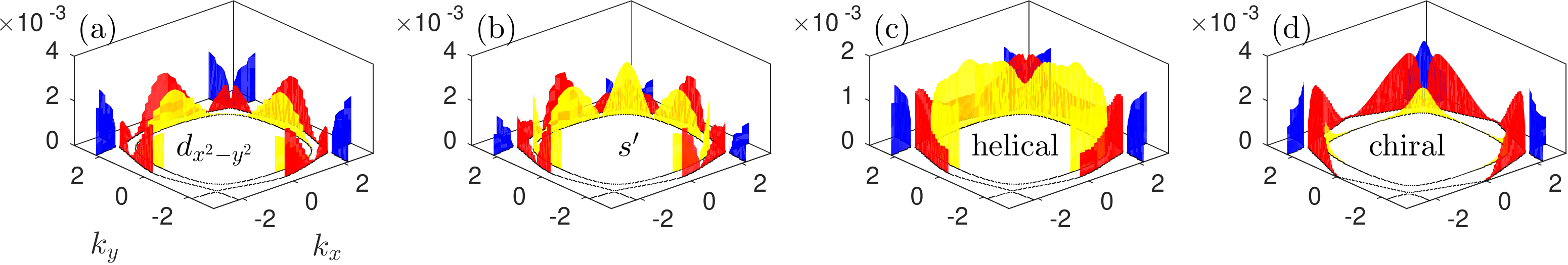}
   	\includegraphics[angle=0,width=.95\linewidth]{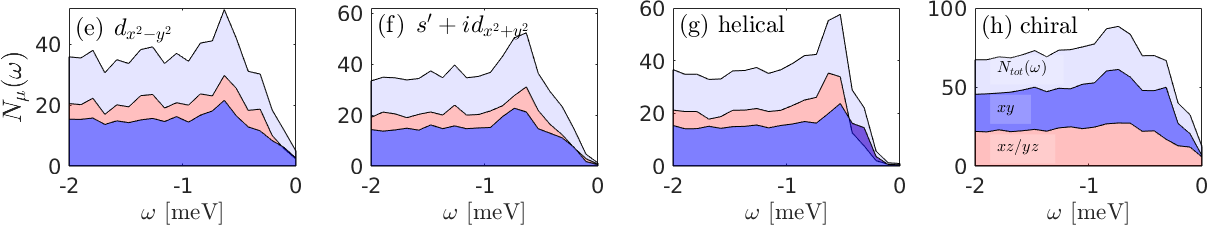}
\caption{Spectral gap $\Delta_\kv$ for (a) $d_{x^2-y^2}$ ($J/U=0.1$), (b) nodal $s^\prime$ ($J/U=0.2$), (c) helical ($J/U=0.25$) in the case of $\mu_{xy}=109$ meV, $\lsoc=35$ meV and $U=120$ meV, see white stars in Fig.~\ref{fig:LGE_Cobo}(b). (d) Spectral gap for the chiral solution with $\mu_{xy}=134$ meV, $\lsoc=35$ meV, $U=50$ meV and $J/U=0.25$. The band character of the gap is indicated by colors; $\alpha$ (blue), $\beta$ (yellow) and $\gamma$ (red). The band character generally corresponds directly to the orbital character, with the exception of the Fermi surface regions close to the zone diagonals as visualized in the Fermi surface insets in Fig.~\ref{fig:LGE_Cobo} (b,d). (e-h) Density of states for $\Delta_{\rm max}=1$ meV. In panels (e,g,h) we invoke the gap structure in (a,c,d) while (f) shows $N_\mu(\omega)$ for a TRSB superconductor constructed by the complex superposition of the two even-parity solutions $d_{x^2-y^2}$ and $s^\prime$.}
\label{fig:gapstruc}
 \end{figure*}

The leading and sub-leading superconducting instabilities are determined from the linearized gap equation
\begin{eqnarray}
  -\int_{FS} d \kv_f^\prime \frac{1}{v(\kv_f^\prime)} \Gamma_{l,l'}(\kv_f,\kv_f^\prime)\Delta_{l'}(\kv_f^\prime)=\lambda \Delta_l(\kv_f),
  \label{eq:LGE}
\end{eqnarray}
where
$    \Delta_l(n,\kv)=\frac{1}{2}\sum_{n^\prime, \kv^\prime,l'}\Gamma_{l,l'}(n,\kv;n',\kv^\prime)\langle \Psi_{l'}(n',\kv')\rangle. $
The integration in Eq.~(\ref{eq:LGE}) includes momenta at the Fermi surface of the three bands with $n$ uniquely defined by $\kv_f$ and $v(\kv_f)$ is the Fermi velocity at $\kv_f$. The eigenvector $\Delta_l(\kv_f)$ corresponding to the largest eigenvalue $\lambda$ displays the structure of the leading superconducting instability.

The solutions to Eq.~(\ref{eq:LGE}) are classified by even parity states, $\Delta_0(\kv)$ with the possible symmetries $\{s,d_{x^2-y^2},d_{xy},g\}$ and odd parity states, i.e.  helical states (four possible superpositions of $\Delta_x(\kv)$ and $\Delta_y(\kv)$) and a chiral solution, $\Delta_z(\kv)$. Here, $\left\{\Delta_x(\kv),\Delta_y(\kv), \Delta_z(\kv)\right\}$ denote the components of the vector ${\bf d(\kv)}$~\cite{SigristUeda} in the {\it pseudospin} space. In our approach, the $x$ and $y$ components are degenerate, due to a lack of hybridization between the $xz$ and $yz$ orbitals. Therefore, all four helical states are degenerate and leaves open the possibility of complex superpositions of the type $\Delta_x+i\Delta_y$, which are non-unitary pair states breaking time-reversal symmetry (TRS).

In Fig.~\ref{fig:LGE_Cobo}(a-d), we show the longitudinal ($zz$) component of the {\it spin} susceptibility and the leading superconducting instabilities as a function of SOC and Hund's coupling $J$ for two different values of $\mu_{xy}=109$, $134$ meV, to expose the effect of van Hove proximity. The Fermi surface in each case is shown in the insets of Fig.\ref{fig:LGE_Cobo}(b,d). The change in $\mu_{xy}$ has a strong effect on the physical susceptibilities, as shown in Fig.~\ref{fig:LGE_Cobo}(a,c) where we plot $\chi_{RPA}^{zz}(\bf{q})$.
For the band farthest from the van Hove point, we observe two prominent nesting vectors, which are approximately given by
$\Qv_1=(2\pi/3,2\pi/3)$ and $\Qv_3=(\pi,2\pi/3)$, see Fig.~\ref{fig:LGE_Cobo}(a). 
The vector $\Qv_1$ arises from the nesting of the 1D-like $xz/yz$ bands, see SM~\cite{suppl}, and has been extensively reported by neutron scattering~\cite{Sidis99,Braden04,Iida11}.
Furthermore, a factor two enhancement of the out-of-plane susceptibility compared to the in-plane susceptibility has been reported at this nesting vector~\cite{Braden04}. Our calculations also give a spin anisotropy at $\Qv_1$ with a magnitude that depends on both SOC, interaction parameters and the band structure, see SM~\cite{suppl}. As shown in Fig. \ref{fig:LGE_Cobo},  the regime where the spin susceptibility is dominated by $\Qv_1$ and $\Qv_3$ results in  mainly even-parity solutions, which are both nodal, $s^\prime$ or $d_{x^2-y^2}$. A helical odd-parity pseudo-spin triplet solution is, however, favored in the regime of large SOC and Hund's coupling $J$, as seen in Fig.~\ref{fig:LGE_Cobo}(b). We stress that for obtaining the results in Fig.~\ref{fig:LGE_Cobo}, it is crucial to properly include SOC both in the band structure {\it and} in the pairing kernel, see SM~\cite{suppl}. Experimentally, the spin anisotropy observed by neutron scattering persist to 300 K~\cite{Braden04} and photo-emission fitting gives a value of $\lambda_{soc}=32$ meV~\cite{Zabolotnyy13}. The Hund's coupling is estimated to be $J/U\simeq 0.1$~\cite{Vaugier12}.
 
Only Fermi surfaces with a $\gamma$-band very close to the van Hove point produce a significant quasi-ferromagnetic signal $\Qv_2$ originating mainly from intra-orbital $xy$ nesting, see Fig.~\ref{fig:LGE_Cobo}(c). At large values of $\lsoc$, chiral pseudo-spin triplet superconductivity emerges as shown in Fig.~\ref{fig:LGE_Cobo}(d). However, when $\Qv_2$ is less pronounced in better agreement with neutron experiments, the chiral state is entirely absent as a leading instability. For further parameter-dependence of the leading superconducting instability, we refer to the SM~\cite{suppl}.

We note that a similar spin-fluctuation based approach was recently employed in Ref.~\onlinecite{Zhang18}, focusing on the very weak-coupling regime and small Hund's interaction. In this limit, chiral or helical solutions were found, whereas even-parity solutions dominated the regime of intermediate coupling strengths. One of our main findings, however, is that a helical state becomes again dominant for the larger values of the Hund's coupling and sizeable SOC, see Fig.\ref{fig:LGE_Cobo}(b). In addition, the chiral state occurs only in regimes where the spin fluctuations appear inconsistent with available neutron scattering data.

The Knight shift provides a way to distinguish between even and odd-parity solutions found in Fig.~\ref{fig:LGE_Cobo}(b,d). 
We address the Knight shift by a calculation of the uniform spin susceptibility in the superconducting state, $\chi^\prime_{0}(\qv=0,\omega=0)$ in four different gap scenarios; $d_{x^2-y^2}$, $s^\prime + i d_{x^2-y^2}$, helical and chiral superconductivity. 

If SOC was negligible, we would expect the Knight shift of the even-parity superconductors to be completely suppressed in all spin channels for $T\to 0$~\cite{Yosida} with exponential suppression for a full gap ($s$-wave) and linear suppression for a nodal gap.
 As seen in Fig.~\ref{fig:LGE_Cobo}(e,f), the even-parity solutions do exhibit suppression in all spin channels, but more pronounced for the in-plane field directions, $xx/yy$.
The simple expectation for singlet superconductors breaks down because a pseudo-spin singlet solution contains both electronic spin singlet {\it and} triplet character. To illustrate this point more clearly, we show in SM~\cite{suppl} how a conventional $s$-wave superconductor acquires a residual Knight shift at $T=0$ as an effect of SOC.
 The properties of helical and chiral solutions, however, remain largely as expected from the $\lsoc=0$ case: 
The helical superconductor exhibits a partial Knight shift suppression for in-plane fields and is insensitive to out-of-plane fields, see Fig.~\ref{fig:LGE_Cobo}(g).
For the chiral state shown in Fig.~\ref{fig:LGE_Cobo}(h), the Knight shift is unaffected by in-plane fields and suppressed by out-of-plane magnetic fields, but full suppression is prevented by SOC~\cite{Wang19}.

Relating to the newest NMR results~\cite{Pustogow19}, our calculations reveal that the superconducting ground state in \sruo~is consistent  either with an even-parity pseudo-spin singlet  or a helical pseudo-spin triplet pair state. Future NMR measurements for out-of-plane fields should be able to distinguish between these cases: the helical solution should exhibit no suppression, while the even-parity solution should display a clear suppression. Finally, we note that a possible non-unitary TRSB state of the type $\Delta_x+i\Delta_y$ would display the same Knight shift as the helical solution.

Turning to the spectral properties of the various superconducting states found above, an outstanding experimental puzzle is the experimental observation of nodes (or near-nodes) in the density of states (DOS) ~\cite{Hassinger17,Suzuki02,Ishida00,Bonalde00,Deguchi04,Suderow98}. For the details of the DOS calculations we refer to the SM section~\cite{suppl}. The $d_{x^2-y^2}$ solution found in Fig.~\ref{fig:LGE_Cobo}(b) has symmetry-imposed line nodes, with a gap that rises very steeply away from the zone diagonals, as shown in Fig.~\ref{fig:gapstruc} (a). The nodes give rise to the characteristic V-shaped DOS at the Fermi level, as shown in Fig.~\ref{fig:gapstruc}(e). The $s^\prime$ solution, which appears to be very prominent in a large region of phase space also exhibits nodes, see Fig.~\ref{fig:gapstruc}(b), but in general the nodes do not coincide with the nodes of $d_{x^2-y^2}$-wave. However, the $\beta$-pocket shows a suppressed $d_{x^2-y^2}$ gap in the region where the $s^\prime$ solution has nodes. Therefore, the  TRSB solution of the type $s^\prime + i d_{x^2-y^2}$ will exhibit near-nodal behavior with a small DOS close to the Fermi level,
as seen in Fig.~\ref{fig:gapstruc} (f). 
The helical state gives rise to a more uniform spectral gap, see Fig.~\ref{fig:gapstruc}(c), with near-nodal behavior only at the $\alpha$ pockets at the zone diagonals. Thus, in this case, we find a more complete suppression of the DOS at the smallest energies, see Fig.~\ref{fig:gapstruc}(g). Finally, for the chiral solution, only segments of the Fermi surface which are predominantly of $xy$ orbital character, display a large gap, as can be deduced by comparing the spectral gap of Fig.~\ref{fig:gapstruc}(d) with the orbital character of the Fermi surface displayed in the inset of Fig.~\ref{fig:LGE_Cobo}(d). Parts of the Fermi surface which are of $xz/yz$ character exhibit almost no gap, and thus there remains a large number of electronic states close to the Fermi surface as evident from Fig.~\ref{fig:gapstruc} (h). We note that this appears to agree with the findings of the recent work by Wang {\it et al.}~\cite{Wang19}, where a chiral solution was found to have low-lying states. The chiral state, however, appears to be ruled out by the recent NMR results~\cite{Pustogow19}.

\begin{figure}[t]
 \centering
   	\includegraphics[angle=0,width=\linewidth]{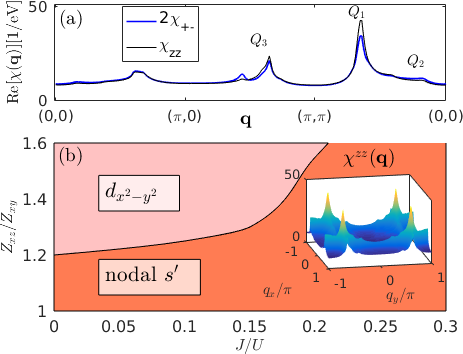}
\caption{(a) $\chi^{zz}({\bf q})/\chi^{+-}({\bf q})$ along the path $(0,0)-(\pi,0)-(\pi,\pi)-(0,0)$ in the case of $Z_{xz}/Z_{xy}=1.6$ for a band with $\mu=90$ meV, $\mu_{xy}=128$ meV and $\lambda_{soc}=35$ meV. The signal at $\Qv_1$ dominates and exhibits a spin anisotropy in rough agreement with experiments \cite{Braden04}.(b) Leading superconducting instability for $\mu,\mu_{xy}$ and $\lsoc$ as in (a) as a function of quasi-particle weight anisotropy $Z_{xz}/Z_{xy}$ and $J$. The inset shows $\chi^{zz}({\bf q})$.}
\label{fig:OrbSelect}
 \end{figure}

In \sruo~significant mass renormalizations have been identified from DMFT originating from the proximity of the van Hove singularity~\cite{Mravlje11} and Hund's coupling, driving the effective mass of the $xy$ orbital larger than the effective mass of $xz/yz$ orbitals.  To investigate how this changes the gap solutions, we apply the same approach as in Refs.~\cite{Kreisel17,Sprau17}. Thus, the bare electronic operator is modified by $c_{\kv,\mu,s} \to \sqrt{Z_\mu}c_{\kv,\mu,s}$ and a difference in quasi-particle weights between the $xy$ orbital and the $xz/yz$ orbitals is imposed by $Z_{xz}=Z_{yz}>Z_{xy}$.
The quasi-particle weights dress the susceptibility\cite{Kreisel17}
\begin{equation}
   \left[ \tilde \chi_0 \right]^{\io_1,\io_2}_{\io_3,\io_4} \rightarrow \sqrt{Z_{\mu_1}} \sqrt{Z_{\mu_2}} \sqrt{Z_{\mu_3}} \sqrt{Z_{\mu_4}} \left[ \chi_0 \right]^{\io_1,\io_2}_{\io_3,\io_4}, 
\end{equation}
and the interaction Hamiltonian Eq.~(\ref{eq:Heff}).
In Fig.~\ref{fig:OrbSelect}(a) we show how the orbital-selective quasi-particle weights lead to improved agreement with the spin susceptibility as measured be neutron scattering\cite{Sidis99,Braden04,Braden02,Iida11}.
For example, the signal at $\Qv_3$ in Fig.~\ref{fig:LGE_Cobo}(a) which
originates  from inter-band nesting between the $xy$ orbital and the
$xz/yz$ bands has been reported by neutron scattering only in Ref.\cite{Iida11}, interpreted as a ridge of the $\Qv_1$ peak with weaker intensity.

A suppression of the response at $\Qv_3$ as well as $\Qv_2$ is observed when we calculate the spin response in the case of stronger mass enhancement of the $xy$ orbital compared to the $xz/yz$ orbitals~\cite{Mravlje11} ($Z_{xz}/Z_{xy}>1$). This scenario leaves the spin anisotropic response at $\Qv_1$ the main magnetic feature of our calculation and provides a route to closer agreement with neutron scattering observations. In this approach, the linearized gap equation results in either nodal $s^\prime$ or $d_{x^2-y^2}$ solutions, and a notable absence of odd-parity pair states, as shown in the phase diagram Fig.~\ref{fig:OrbSelect} (b). The large boundary between the two solutions points to the possibility of a $s^\prime+id_{x^2-y^2}$ gap structure which could reconcile the properties of 1) a decrease in Knight shift for in-plane fields at $T<T_c$, 2) nodal low-energy electronic states available for transport, and 3) signatures of TRSB \cite{Luke1998,Kapitulnik09}\footnote{ We note that our 2D calculations are not 
capable of capturing pairing instabilities that pair electrons between two layers, as proposed in Ref. \onlinecite{Pustogow19}.}.

In summary we have provided a timely theoretical study of the leading superconducting instabilities in \sruo. We have discussed their spectral and magnetic properties and focused on recent neutron scattering and Knight shift measurements, which seem inconsistent with chiral triplet pairing and point to other preferred pair states for this material. Several possibilities are discussed, including a rare helical triplet state and more prevalent even-parity pair states which, as we have shown, can be distinguished by future experiments.      

The authors are grateful for illuminating discussions with S. Brown, P. Kotetes, A. Kreisel, S. Mukherjee, S. Raghu and  P. Steffens.
A.T.R., D.D.S., and B.M.A. acknowledge support from the Carlsberg Foundation.  P.J.H. was supported by  the U.S. Dept. of Energy  under Grant No. DE-FG02-
05ER46236.
\nocite{*}
 \bibliography{bibliography_sr2ruo4}
\begin{widetext}
 
\end{widetext}

\section{Supplementary Material: \\
Knight Shift and Leading Superconducting Instability From Spin Fluctuations in \sruo }

This supplementary material provides additional details about the spin susceptibility and the main orbital contributions to the spin response. In addition, we provide information about
leading and sub-leading instabilities of the superconducting order in \sruo~, the sensitivity to the electronic bands, spin-orbit coupling and interaction strengths.  Furthermore, we present the derivation the effective pairing interaction in the presence of spin-orbit coupling, and  elaborate on the details of the linearized gap equation, the Knight shift and the density of states calculations.
\begin{figure}[t]
 \centering
   	\includegraphics[angle=0,width=.9\linewidth]{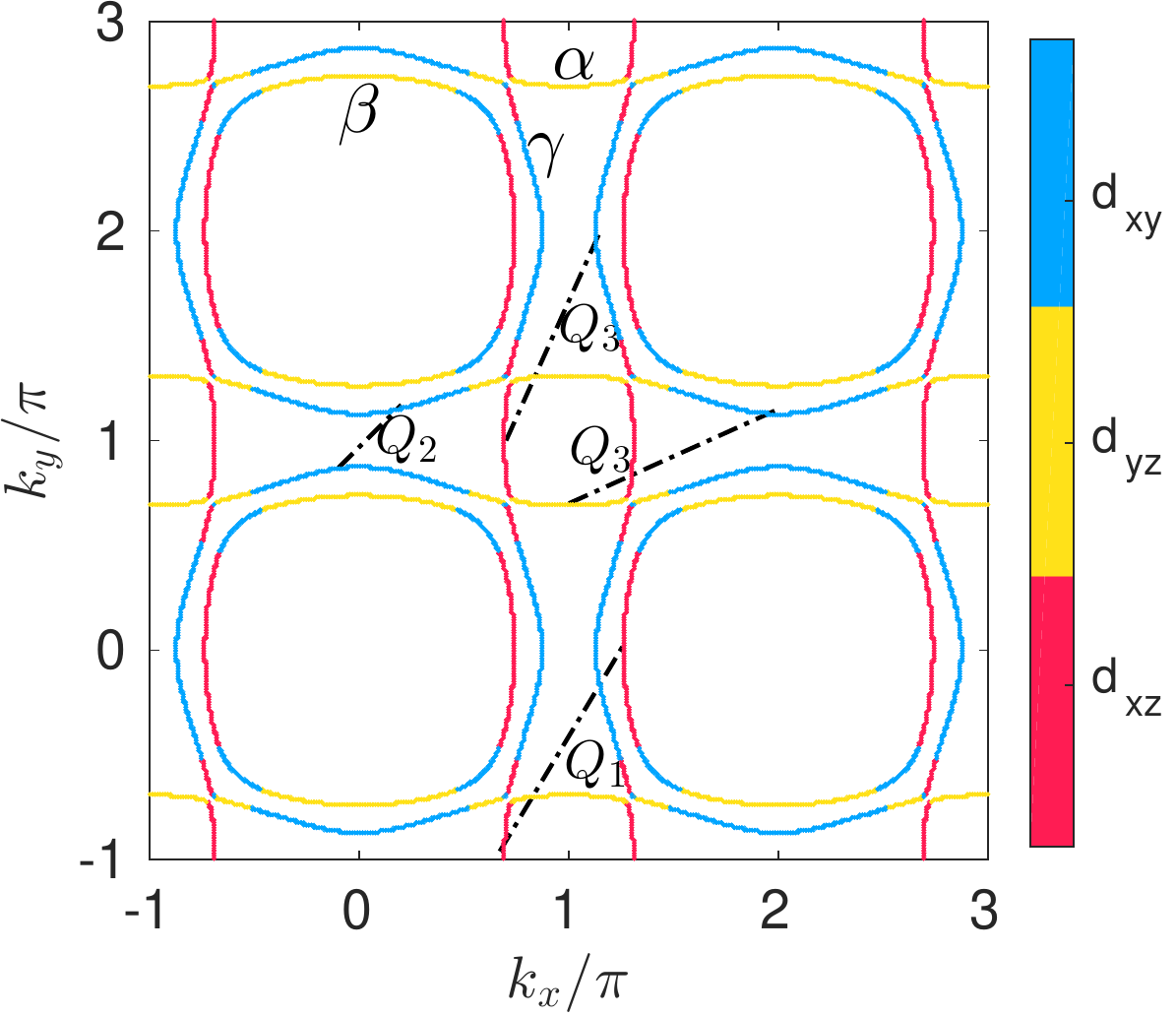}
\caption{Fermi surface in the extended zone scheme displaying the dominating orbital content, $d_{xz}$ (red), $d_{yz}$ (yellow), and $d_{xy}$ (blue) at the $(\pi,\pi)$-centered $\alpha$-pocket and the $\beta$ and $\gamma$-pockets centered at $(0,0)$. We set $\mu=\mu_{xy}=109$ meV, $\lambda_{soc}=35$ meV and the hopping constants are as stated in the main text. The three nesting vectors $\Qv_1$, $\Qv_2$ and $\Qv_3$ are depicted by dashed-dotted lines. Note that the nature of the nesting at $\Qv_1$ and $\Qv_2$ is intra-orbital while the nature of the nesting at $\Qv_3$ is inter-orbital.
}
\label{fig:nesting}
 \end{figure}
~

\section{Spin susceptibility and nesting vectors}
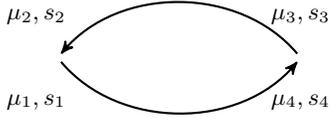
\begin{figure}[t]
\begin{center}
\begin{tikzpicture}[node distance=1cm, auto,]
 \node[] (markets) {};
 \node[right=3cm of markets] (market) {}
 (market.south) edge[pil, bend right=50] (markets.south);
 \node[right=2cm of markets] (0) {} 
 (markets.south) edge[pil, bend right=50] (market.south);
\node[above=.1cm of market] (nu2w) {$\mu_3, s_3$};
\node[below=0.7cm of nu2w] (nu1w) {$\mu_4, s_4$};
\node[left=2.5cm of nu2w] (nu3w) {$\mu_2, s_2$};
\node[below=0.7cm of nu3w] (nu3w) {$\mu_1, s_1$};
\end{tikzpicture}
\end{center}
\caption{The generalized bare spin susceptibility, $[\chi_0]^{\mu s_1, \mu s_2}_{\mu s_3, \mu s_4}(\qv)$. Main contributions to $\chi_0$ are described in Table 1. For non spin-flip processes, $s_1=s_4=s'$ and $s_2=s_3=s$, and  the orbital indices satisfy $\mu_1=\mu_4=\mu'$  and $\mu_2=\mu_3=\mu$. For the main spin flip processes, we have e.g. $\mu_1=\mu'$ and $\mu_2=\mu_3=\mu_4=\mu$ with the corresponding spins $s_1=\overline s_4$.  }
\label{fig:chi0}
\end{figure}

In this section we investigate the components of the spin susceptibility in order to determine the physical nature of the dominating processes.
Inspection of the components of the bare susceptibility provides us all the largest contributions to the generalized susceptibility which is given by:
\begin{widetext}
\begin{eqnarray}
 [\chi_0]^{\mu_1 s_1,\mu_2s_2}_{\mu_3s_3,\mu_4s_4}(\qv,i\omega_n)&=&\frac{1}{N}\int_0^\gamma d \tau e^{i \omega_n \tau} \sum_{\kv,\kv'} \langle T_\tau c^\dagger_{\kv- \qv \mu_1s_1} (\tau)
c_{\kv \mu_2s_2}  (\tau) c^\dagger_{\kv'+ \qv \mu_3s_3} (0)  c_{\kv' \mu_4s_4} (0) \rangle_0.
 \end{eqnarray}
In the normal state, this becomes
 \begin{eqnarray}
 [\chi_0]^{\mu_1s_1,\mu_2s_2}_{\mu_3s_3,\mu_4s_4}(\qv,i\omega_n)&=&
-\frac{1}{N}\sum_\kv \sum_{n_1,n_2}[M_{n_1,n_2}(\kv,\qv)]^{\mu_1\spin_1,\mu_2\spin_2}_{\mu_3\spin_3,\mu_4\spin_4}
 \frac{f(\xi_{\kv-\qv, n_1 ,\spin_1})-f(\xi_{\kv ,n_2 ,\spin_2})}{i\omega_n+\xi_{\kv-\qv, n_1, \spin_1}-\xi_{\kv, n_2 ,\spin_2}},
\label{eq:BareSus}
 \end{eqnarray}
with
\begin{eqnarray}
 [M_{n_1,n_2}(\kv,\qv)]^{\mu_1\spin_1,\mu_2\spin_2}_{\mu_3\spin_3,\mu_4\spin_4}=[u_{n_1\spin_1}^{\mu_1 s_1}(\kv-\qv)]^*  [u_{n_2\spin_3}^{\mu_3 s_3} (\kv)]^* u_{n_2\spin_2}^{\mu_2 s_2} (\kv)u_{n_1\spin_4}^{\mu_4 s_4} (\kv-\qv),
\end{eqnarray}
\end{widetext}
where $u_{n\spin}^{\mu s}(\kv)$ is the eigenvector of the transformation from orbital and electronic spin basis $(\mu,s)$ to band and pseudo-spin basis $(n,\sigma)$.
In Table~\ref{tab1}, we list the three prominent nesting vectors, $\Qv_1,\Qv_2,\Qv_3$, and how they are related to orbital- and spin degrees of freedom. The corresponding bare susceptibility diagram is drawn in Fig.~\ref{fig:chi0}.
The exact wave vectors will be band dependent, but the orbital origin and spin character of the main contributions are band independent. 
The physical processes responsible for the response at $\Qv_1$ and $\Qv_2$ are intra-orbital non-spin flip processes.
The processes responsible for the response with wave vector $\Qv_3$ are inter-orbital scatterings, which are either spin-conserving or spin-flipping.
All main contributions that involve the $xy$ orbital are very sensitive to the proximity of the van Hove singularity of the $\gamma$ band at $(\pm\pi,0)/(0,\pm\pi)$ to the Fermi surface. This means that a shift of $\mu_{xy}$  modifies the response at $\Qv_2$ and $\Qv_3$.

\section{Effect of interactions}

\begin{figure*}[t]
 \centering
   	\includegraphics[angle=0,width=.32\linewidth]{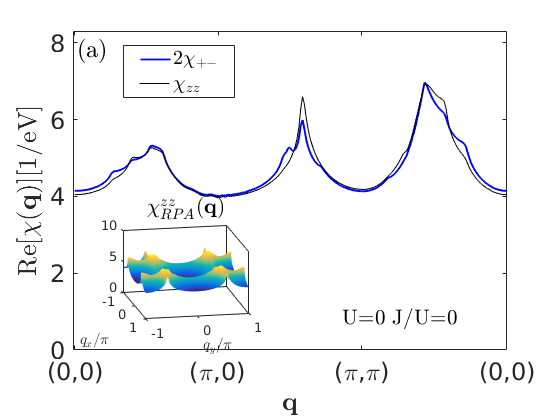}
   	   	\includegraphics[angle=0,width=.3\linewidth]{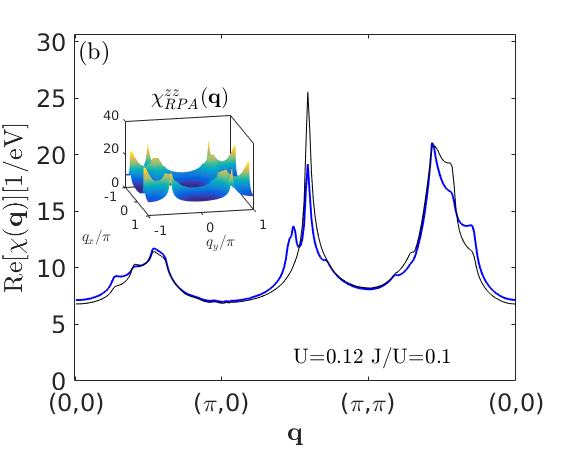}
   	   	   	\includegraphics[angle=0,width=.32\linewidth]{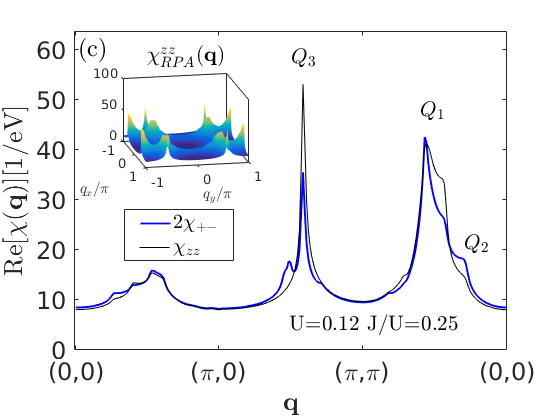}
\caption{Longitudinal and transverse susceptibility as defined in Eq.~(\ref{eq:chiphys}) for $\sigma^a=\sigma^b=\sigma^z$ and $\sigma^a=\sigma^+$ and $\sigma^b=\sigma^-$, respectively. (a) displays the bare susceptibility, and (b,c) show the interacting case with $U=120$ meV and $J/U=0.1,0.2$, respectively. In all cases $\lambda_{soc}=35$ meV and the chemical potential is $\mu=\mu_{xy}=109$ meV corresponding to the case shown in Fig. 1 (a,b) of the main text.}
\label{fig:chiphys}
\end{figure*}

The effect of interactions is taken into account by the random-phase approximation (RPA):
\begin{eqnarray}
\Big[\chi\Big]^{\io_1 , \io_2 }_{\io_3,\io_4 } (\qv)&=&\Big[\frac{1}{1-\chi_0U}\chi_0\Big]^{\io_1 \io_2}_{\io_3 \io_4}(\qv) 
\label{eq:chiRPA}
\end{eqnarray}
where $\chi_0$ and $U$ are $36\times 36$ matrices and we introduce a joint index $\tilde \mu=(\mu,s)$ for orbital and electronic spin. In the matrix multiplication above we keep in mind that the matrices in general do not commute. Therefore, the RPA construction is less transparent than in a simple one-band calculation, where the effect of interactions within RPA simply amounts to an enhancement of the bare susceptibility and the superconducting instabilities from a spin-fluctuation mechanism can be related to the bare spin susceptibility in a relatively straightforward fashion (see e.g. A. T. R\o mer {\it et al.}, Phys. Rev. B, {\bf 92}, 104505 (2015)). Additional features arise due to the presence of sizable spin-orbit coupling, which is responsible for the spin anisotropy between in-plane $(xx,yy)$ and out-of-plane $(zz)$ components of the susceptibility. 
To illustrate these points, we show in Fig.~\ref{fig:chiphys}, the longitudinal and transverse susceptibility for a realistic spin-orbit coupling of $\lambda_{soc}=35$ meV as a function of increasing interaction parameters $U$ and $J$.

\begin{table}[t]
\begin{tabular}{|c|c|c|c|}
\hline
\hline
\multicolumn{4}{c}{ Spin-preserving  processes} \\
\hline
Wave vector & \multicolumn{2}{l}{Orbital character}  & Spin character \\
\hline
 & $\mu$ & $\mu'$ & $(s_1,s_2,s_3,s_4)$  \\
\hline
$\Qv_1$ & $xz$ & $xz$ & $(s,s,s,s)/(\overline{s},s,s,\overline{s})$ \\
({\it intraorbital }) & $yz$ & $yz$ & $(s,s,s,s)/(\overline{s},s,s,\overline{s})$ \\
\hline
$\Qv_2$ & $xy$ & $xy$ &  $(s,s,s,s)/(\overline{s},s,s,\overline{s})$\\
({\it intraorbital}) & & & \\
\hline
$\Qv_3$ & $xz$ & $xy$ &  $(s,s,s,s)/(\overline{s},s,s,\overline{s})$\\
({\it interorbital}) & $yz$ & $xy$ &  $(s,s,s,s)/(\overline{s},s,s,\overline{s})$\\
\hline
\multicolumn{4}{c}{ Spin-flip processes} \\
\hline
 & $\mu$ & $\mu'$ & $(s_1,s_2,s_3,s_4)$  \\
\hline
$\Qv_3$ & $xy$ & $yz$ & $(s,s,s,\overline{s})$  \\
({\it interorbital}) & $xy$ & $yz$ & $(\overline{s},s,s,s)$  \\
 & $yz$ & $xy$ & $(s,s,s,\overline{s})$  \\
 & $yz$ & $xy$ & $(\overline{s},s,s,s)$ \\
\hline
\hline
\end{tabular}
\caption{Main contributions to the bare spin susceptibility $[\chi_0]^{\mu_1 s_1, \mu_2 s_2}_{\mu_3 s_3, \mu_4 s_4}(\qv)$ depicted in Fig.~\ref{fig:chi0}. For spin-preserving processes $\mu_1=\mu_4$ and $\mu_2=\mu_3$. For spin-flip processes (linear in $\lambda_{soc}$), one orbital index differs from the remaining.}
\label{tab1}
\end{table}

The physical susceptibility is derived from the generalized susceptibility by
\begin{eqnarray}
 \Big[\chi^{ab}(\qv,i\omega_n)\Big]&=& \frac{1}{N}\int_0^\gamma d \tau e^{i \omega_n \tau} \sum_{\mu,\nu} \langle T_\tau S_{\mu}^a (-\qv,\tau) S_{\nu}^b (\qv,0) \rangle \nonumber \\
 &=& \frac{1}{4}\sum_{\mu,\nu,\{s\}} \spin_{s_1s_2}^a \spin_{s_3,s_4}^b [\chi_0]^{ \mu, s_1; \mu,s_2}_{\nu, s_3;\nu s_4}(\qv,i\omega_n), \nonumber\\
 \label{eq:chiphys}
\end{eqnarray}
where the matrices $\spin^a/\spin^b$ are the Pauli matrices.

\begin{figure}[b]
 \centering
   	\includegraphics[angle=0,width=\linewidth]{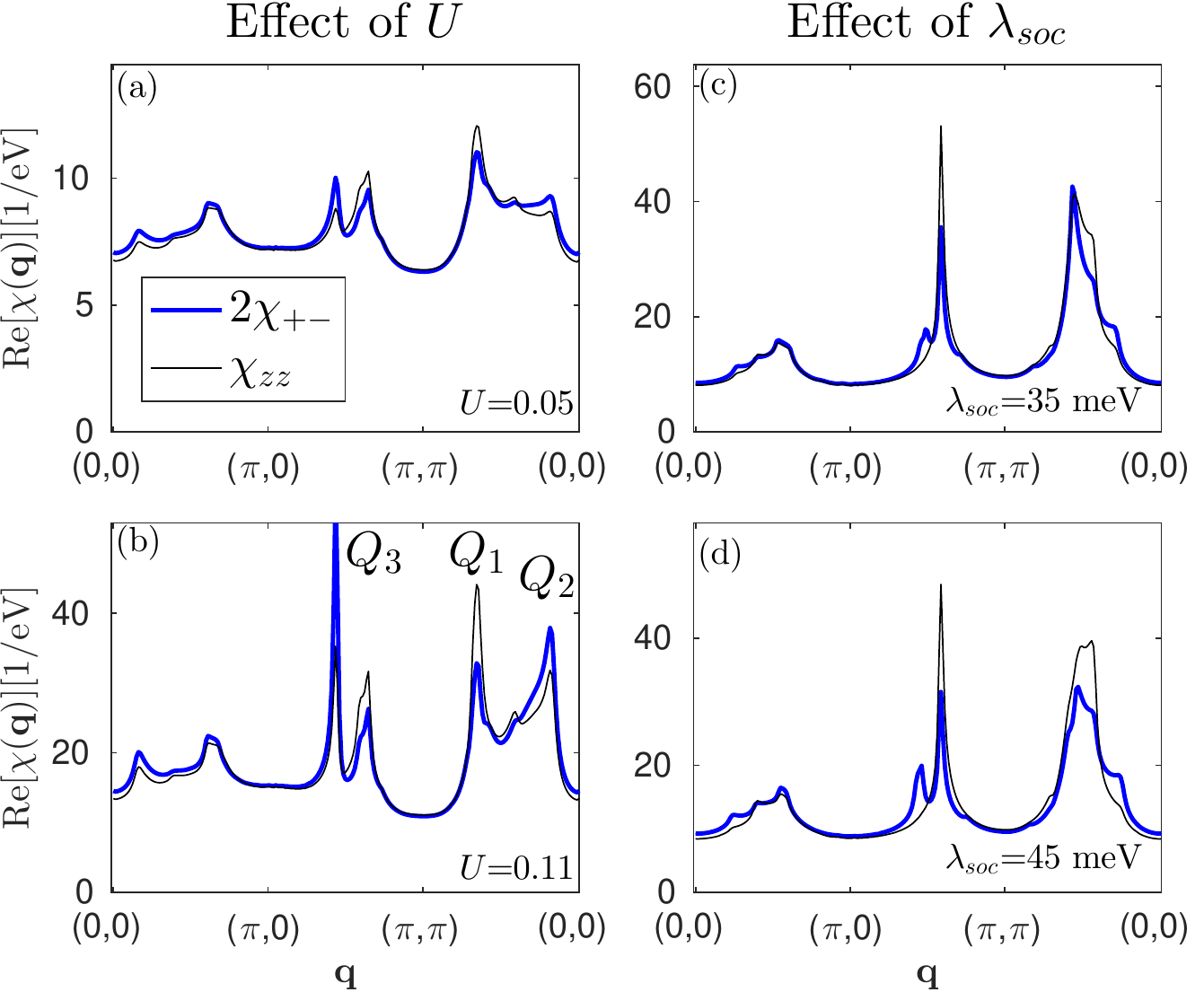}
\caption{(a,b) Spin susceptibilities for band parameters $\mu=90$ meV, $\mu_{xy}=128$ meV and $\lambda_{soc}=35 $ meV (hopping constants are specified in the main text).
An increase of $U$ gives rise to larger spin anisotropy (in both cases we set $J/U=0.25$). (c,d) Spin susceptibilities for band parameters $\mu=\mu_{xy}=109$ meV. An increase in $\lambda_{soc}$ from 35 meV to 45 meV enhances the spin anisotropy at $\Qv_1$. In both cases we set $U=120$ meV and $J/U=0.25$.}
\label{fig:effectofinteractions}
\end{figure}

In Fig.~\ref{fig:chiphys} we display the spin susceptibilities relevant for Fig. 1 (a,b) of the main text. We note how the spin susceptibilities are enhanced by $U$ and $J$, and the spin anisotropy becomes more pronounced as $J$ increases. For the band shown in Fig.~\ref{fig:chiphys}, both signals at $\Qv_1$ and $\Qv_3$ are strong and the spin anisotropy appearing as a shoulder formation in $\chi_{zz}$ at $\Qv_1$ is less significant than reported by neutron scattering experiments~\cite{Braden04}.
To further display the sensitivity of the spin anisotropy to band structure details, $U$ and $\lambda_{soc}$, we show in Fig.~\ref{fig:effectofinteractions} the spin susceptibilities for two different values of $\mu$ and $\mu_{xy}$. In the first case, Fig.~\ref{fig:effectofinteractions}(a,b), the structure at the position $\Qv_1$ is sharp and shows a clear spin anisotropy, especially at larger values of $U$. In the second case show in Fig.~\ref{fig:effectofinteractions}(c,d), the signal at $\Qv_1$ is broader with only small spin anisotropy for moderate values of $\lambda_{soc}$, which is however enhanced for larger values of $\lambda_{soc}$, see Fig. ~\ref{fig:effectofinteractions} (d). The spin susceptibilities shown in  Fig. ~\ref{fig:effectofinteractions}(c,d) fall in the regime of helical superconductivity of Fig. 1 (b) of the main text, while the spin susceptibilities shown in  Fig. ~\ref{fig:effectofinteractions}(a,b) support $s^\prime$-wave superconductivity.

\begin{figure*}[t]
 \centering
   	\includegraphics[angle=0,width=.32\linewidth]{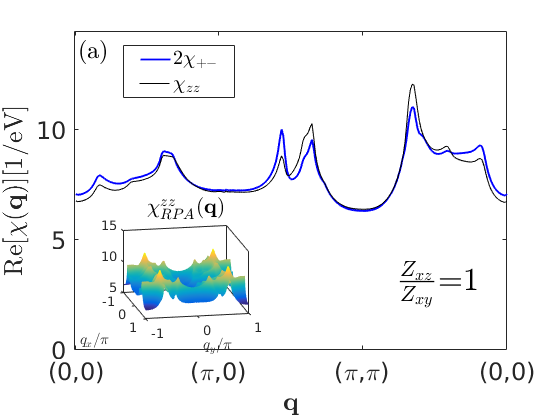}
   	   	\includegraphics[angle=0,width=.32\linewidth]{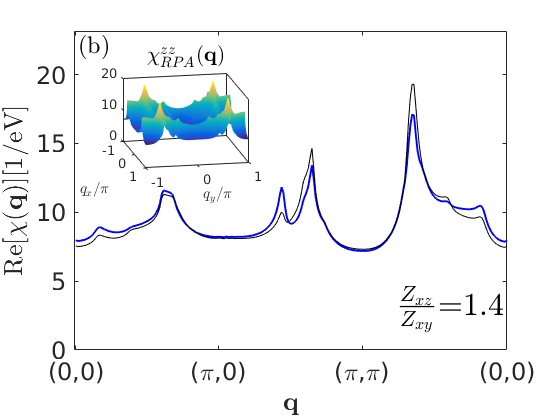}
   	   	   	\includegraphics[angle=0,width=.32\linewidth]{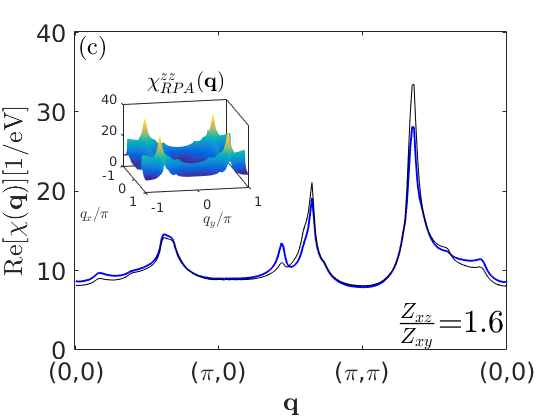}
\caption{Orbital renormalization of longitudinal and transverse susceptibility as defined in Eq.~\ref{eq:chiphys} for $\sigma^a=\sigma^b=\sigma^z$ and $\sigma^a=\sigma^+$ and $\sigma^b=\sigma^-$, respectively for $\mu=90$ meV, $\mu_{xy}=128$ meV, $\lambda_{soc}=35$ meV and interaction parameters $U=50$ meV and $J/U=0.25$. (a) displays the case of no mass renormalization $Z_{xz}=Z_{xy}$ with (b,c) show the renormalization of $\frac{Z_{xz}}{Z_{xy}}=1.4,1.6$, respectively.}
\label{fig:chiphys_OrbSelect}
\end{figure*}

To improve the agreement between our spin susceptibility calculations and the spin susceptibility observed by neutrons, we invoke a phenomenological approach, where the mass renormalizations of the $xz/yz$ and $xy$ orbitals are taken into account, as described in the main text and previously explored in the case of FeSe~\cite{Kreisel17}.
The quasi-particle weights dress the bare electronic operators and thereby the
susceptibility\cite{Kreisel17}
\begin{equation}
   \left[ \tilde \chi_0 \right]^{\io_1,\io_2}_{\io_3,\io_4} \rightarrow \sqrt{Z_{\mu_1}} \sqrt{Z_{\mu_2}} \sqrt{Z_{\mu_3}} \sqrt{Z_{\mu_4}} \left[ \chi_0 \right]^{\io_1,\io_2}_{\io_3,\io_4}, 
\end{equation}
as well as the interaction Hamiltonian. An equivalent formulation is to attach the quasi-particle weights to the bare interaction parameters  $U,J,U'$ and $J'$:
\begin{equation}
    \Big[\tilde U\Big]^{\io_1 , \io_2 }_{\io_3,\io_4 }=\sqrt{Z_{\mu_1}}\sqrt{Z_{\mu_2}}\sqrt{Z_{\mu_3}}\sqrt{Z_{\mu_4}}\Big[U\Big]^{\io_1 , \io_2 }_{\io_3,\io_4 }
\end{equation}
For the band parameters $\mu=90$ meV and $\mu_{xy}=128$ meV we obtain as a function of increasing renormalization $Z_{xz}/Z_{xy}$ the susceptibilities shown in Fig.~\ref{fig:chiphys_OrbSelect}. As the ratio $Z_{xz}/Z_{xy}$ increases, the relative strength of the signal at $\Qv_3$ is weakened in agreement with the neutron report Ref.~\cite{Iida11}. The relative strength of the spin anisotropy at $\Qv_1$ around $20-30 \%$ is only weakly affected by orbital renormalization, see Fig.~\ref{fig:chiphys_OrbSelect}. 

\section{Subleading superconducting instabilities }

In Fig. 1 of the main text, we show the phase diagrams of the leading instabilities as a function of $(J,\lambda_{soc})$ for two different bands, which differ by the proximity of the van Hove instability. In the intermediate case, the nodal $s^\prime$ dominates for all $(J,\lambda_{soc})$.
Here, for completeness we show in Fig.~\ref{fig:LGE_Jdependence} the leading and subleading instabilities for a fixed value of $\lambda_{soc}$ in each case of $\mu_{xy}=109,122,134$ meV. While $d_{xy}$ and $g$-wave appears to be suppressed for all cases of $\mu_{xy}$, the nodal $s^\prime$, $d_{x^2-y^2}$ and helical solutions are close in energy when the energy bands are not tuned too close to the van Hove instability by $\mu_{xy}$. Only in the extreme case of $\mu_{xy}=134$ meV do we find that $s^\prime$ and $d_{x^2-y^2}$ become suppressed and the two odd parity solutions become close in energy, and actually degenerate at $J=0$. In the last regime, chiral superconductivity appears at $J>0$.

\section{Derivation of the effective pairing interaction}
In this section we show the details of the derivation of the effective electron-electron interaction in the Cooper channel as given in the main paper Eq.~(4).
The effective pairing interaction by spin-fluctuations in the multi-orbital system with spin-orbit coupling is derived from the rotationally invariant interaction Hamiltonian
\begin{eqnarray}
 \hat H_{int}&=&U\sum_{i,\mu} n_{i\mu\up}n_{i\mu\down}+\frac{U'}{2}\sum_{i,\nu\neq\mu,s} n_{i\mu s}n_{i\nu\overline{s}}\nonumber \\
 &&+\frac{U'-J}{2}\sum_{i,\nu\neq\mu,s} n_{i\mu s}n_{i\nu s} +\frac{J}{2}\sum_{i,\nu\neq\mu,s} c_{i\mu s}^\dagger c_{i\nu\overline{s}}^\dagger c_{i\mu\overline{s}} c_{i\nu s}
\nonumber \\
&& +\frac{J'}{2}\sum_{i,\nu\neq\mu,s} c_{i\mu s}^\dagger c_{i\mu\overline{s}}^\dagger c_{i\nu\overline{s}} c_{i\nu s} + h.c. 
\end{eqnarray}
where $i$ is the site index, $\mu,\nu$ are orbital indices and $s=-\overline{s}$ refers to real electronic spins. As usual, intra- and interorbital Coulomb scattering as well as pairhopping terms are included, and $U^\prime=U-2J$, $J^\prime=J$.
The interaction Hamiltonian restricted to the Cooper channel can be written in an abbreviated form 
\begin{eqnarray}
\hat H_{int}=\sum_{ \kv,\kv' \{\tilde \mu\}}\Big[U\Big]^{\io_1 , \io_2 }_{\io_3,\io_4 } \quad c_{\kv \io_1 }^\dagger  c_{-\kv \io_3 }^\dagger c_{-\kv' \io_2 } c_{\kv' \io_4 } \nonumber \\
 \label{eq:Heffs}
\end{eqnarray}
where we collect the orbital index, $\mu$, and the electronic spin index, $s$, in one common index; $\io:=(\mu,s)$.
\begin{figure}[b]
\centering

\begin{tikzpicture}[node distance=1cm, auto,]
 \node[punkt] (market) {$\Big[V(\kv,\kv')\Big]^{\tilde \mu_1,\tilde \mu_2}_{\tilde \mu_3,\tilde \mu_4}$};
 \node[above=of market] (dummyup) {};
 \node (sum3) at ($(market.east)!0.33!(dummyup.south)$) {};
 \node[left=1.32cm of sum3] (sum1) {};
 \node[below=.52cm of sum1] (sum4) {};
 \node[below=.52cm of sum3] (sum2) {};
 \node[left=of dummyup] (1) {$\kv, \tilde \mu_1$}
    (sum1.south) edge[pil, bend left=0] (1.east);
\node[right=of dummyup] (3) {$-\kv,\tilde \mu_3$}
 (sum3.south) edge[pil,bend left=0] (3.west);  
\node[below=of market] (dummydown) {};
 \node[left=of dummydown] (4) {$\kv', \tilde \mu_4$}
  (4.east) edge[pil, bend left=0] (sum4);
 \node[right=of dummydown] (2) {$-\kv', \tilde \mu_2$}
   (2.west) edge[pil, bend left=0] (sum2);

   (t);
\end{tikzpicture}
\caption{Index labels of the effective pairing vertex.
We have collected the orbital index, $\mu$, and the electronic spin index, $s$, in one common index; $\io:=(\mu,s)$.}
\label{fig:Veff}
\end{figure}
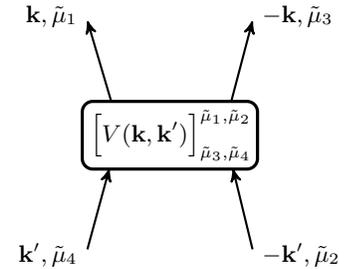
The bare electron-electron interaction, $[U]^{\mu_1 s_1 \mu_2 s_2}_{\mu_3 s_3 \mu_4 s_4}$, is given by
\begin{eqnarray}
 &&\Big[U\Big]^{\mu s \mu  \overline s}_{\mu  \overline s \mu s}=U \qquad \Big[U\Big]^{\nu s \mu  \overline s}_{\mu  \overline s\nu  s}=U' \qquad \Big[U\Big]^{\mu s\nu \overline s}_{\mu \overline s\nu s}=J' \nonumber \\
 &&
 \Big[U\Big]^{\mu s\mu \overline s}_{\nu \overline s\nu s}=J \qquad \Big[U\Big]^{\mu s\nu s}_{\nu s\mu s}=U'-J \nonumber \\
 && \nonumber \\
 &&\Big[U\Big]^{\mu s \mu s}_{\mu  \overline s \mu  \overline s}=-U \quad \Big[U\Big]^{\nu s \nu s}_{\mu  \overline s \mu  \overline s}=-U' 
 \quad \Big[U\Big]^{\mu s\nu s}_{\mu \overline s\nu \overline s}=-J' \nonumber \\
 &&
\Big[U\Big]^{\mu s\nu s}_{\nu \overline s\mu \overline s}=-J \quad \Big[U\Big]^{\mu s\mu s}_{\nu s\nu s}=-U'+J \nonumber \\
 \label{eq:BareU}
\end{eqnarray}
Higher order interactions in $[U]$ are derived diagrammatically from ladder and bubble diagrams. The form of the final interaction Hamiltonian is:
\begin{eqnarray}
\hat H_{int}=\frac{1}{2}\sum_{ \kv,\kv' \{\tilde \mu\}}\Big[V(\kv,\kv')\Big]^{\io_1 , \io_2 }_{\io_3,\io_4 }  \quad c_{\kv \io_1 }^\dagger  c_{-\kv \io_3 }^\dagger c_{-\kv' \io_2 } c_{\kv' \io_4 } \nonumber \\
 \label{eq:Heff_final}
\end{eqnarray}
with the effective interaction $\Big[V(\kv,\kv')\Big]^{\io_1 , \io_2 }_{\io_3,\io_4 }$ shown in Fig.~\ref{fig:Veff}. 
 \begin{figure*}[t]
 \centering
   	\includegraphics[angle=0,width=\linewidth]{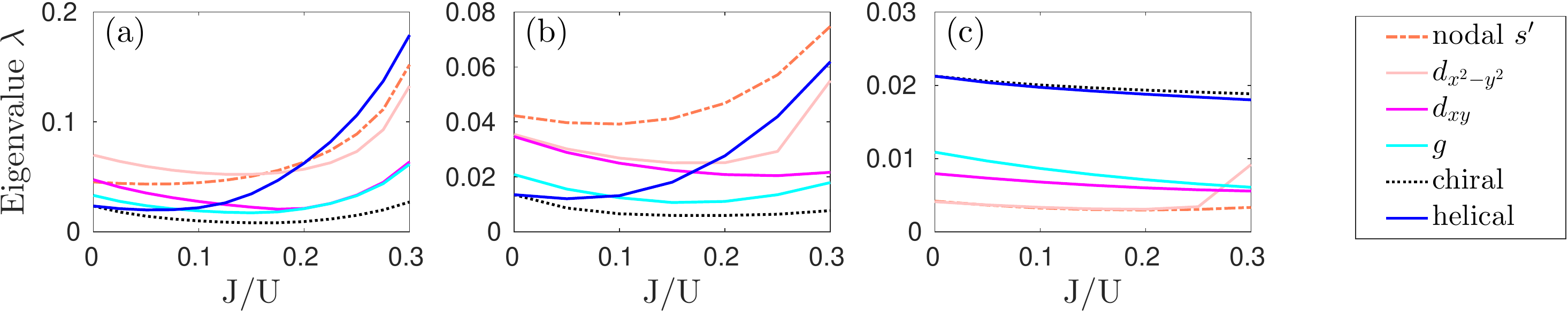}
\caption{Leading and subleading superconducting instability as a function of Hund's coupling $J$ for spin-orbit coupling $\lambda_{soc}=35$ meV for bands of $\mu=109$ meV and (a) $\mu_{xy}=109$ meV, $\lambda_{soc}=35$ meV and $U=120$ meV, (b) $\mu_{xy}=122$ meV, $\lambda_{soc}=30$ meV and $U=100$ meV, and (c) $\mu_{xy}=134$ meV, $\lambda_{soc}=35$ meV and $U=50$ meV. Only the largest eigenvalue of each irreducible representation is depicted, i.e. higher order intermediate instabilities are not shown. We have checked that the order of the solutions is unchanged by moderate changes in $U$.}
\label{fig:LGE_Jdependence}
 \end{figure*}

We sum up all bubble diagrams and ladder diagrams to infinite order in $U$ to obtain the final effective electron-electron interaction. The second order diagrams are shown in Fig.~\ref{fig:bubbleladder}. For the bubbles we obtain the interaction contribution:
\begin{eqnarray}
 \Big[ V_{\rm bub}(\kv,\kv') \Big]^{\io_1\io_2}_{\io_3 \io_4}
&=&-\Big[U\frac{1}{1-\chi_0U}\chi_0U  \Big]^{\io_1\io_4}_{\io_3 \io_2}(\kv-\kv'). \nonumber \\
\end{eqnarray}
The ladder type of diagrams give
\begin{eqnarray}
\Big[ V_{\rm lad} (\kv,\kv')\Big]^{\io_1 \io_2}_{\io_3 \io_4}&=& \Big[U\frac{1}{1-\chi_0U}\chi_0U\Big]^{\io_1 \io_2}_{\io_3 \io_4}(\kv+\kv').\nonumber \\
\end{eqnarray}
The final result for the interaction vertex entering the Hamiltonian in Eq.~(\ref{eq:Heff_final}) is
\begin{eqnarray}
\Big[V(\kv,\kv')\Big]^{\io_1 , \io_2 }_{\io_3,\io_4 } &=&\Big[U\Big]^{\io_1 , \io_2 }_{\io_3,\io_4 }+\Big[U\frac{1}{1-\chi_0U}\chi_0U\Big]^{\io_1 \io_2}_{\io_3 \io_4}(\kv+\kv') \nonumber \\
&&-\Big[U\frac{1}{1-\chi_0U}\chi_0U  \Big]^{\io_1\io_4}_{\io_3 \io_2}(\kv-\kv'),
\end{eqnarray}

where the generalized spin susceptibility  is given by
\begin{eqnarray}
&& [\chi_0]^{\tilde \mu_1,\tilde \mu_2}_{\io_3,\io_4}(\qv,i\omega_n)= \nonumber  \\
&& \frac{1}{N}\int_0^\gamma d \tau e^{i \omega_n \tau} \sum_{\kv,\kv'} \langle T_\tau c^\dagger_{\kv- \qv \io_1} (\tau)
c_{\kv \io_2}  (\tau) c^\dagger_{\kv'+ \qv \io_3}  c_{\kv' \io_4} \rangle_0. \nonumber \\
&&
 \end{eqnarray}
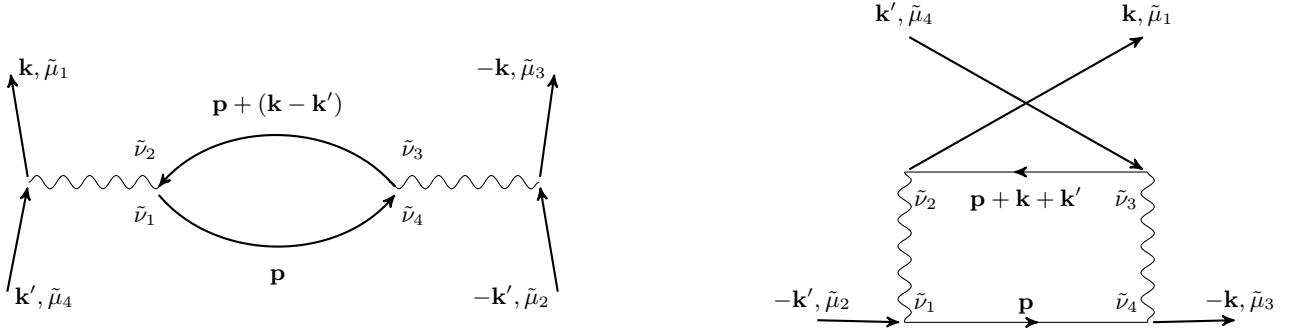
\begin{figure*}[t]
\centering

\begin{tikzpicture}[node distance=1cm, auto,]

 \node[] (markets) {};
 \node[right=3cm of markets] (market) {}
 (market.south) edge[pil, bend right=50] (markets.south);
 \node[right=2cm of markets] (0) {} 
 (markets.south) edge[pil, bend right=50] (market.south);
\node[left=4.8cm of market] (in) {};
\node[left=2.7cm of market] (nu2) {}
(in) edge[photon] (nu2);
\node[above=.2cm of nu2] (nu2wi) {};
\node[left=.05cm of nu2wi] (nu2w) {$\tilde \nu_2$};
\node[below=0.4cm of nu2w] (nu1w) {$\tilde \nu_1$};
\node[left=.01cm of market] (nu3) {};
\node[right=2.cm of nu3] (in2) {}
(in2) edge[photon] (nu3);
\node[above=.2cm of nu3] (nu3wi) {};
\node[right=.05cm of nu3wi] (nu3w) {$\tilde \nu_3$};
\node[below=0.4cm of nu3w] (nu3w) {$\tilde \nu_4$};
\node[above=.8cm of market](ppr) {};
\node[left=.5cm of ppr](pp) {$\pv+(\kv-\kv')$};
\node[below=1.8 of pp](0) {$\pv$};
\node[left=2cm of pp] (mu1b) {};
\node[above=.1cm of mu1b] (mu1) {$\kv,\tilde\mu_1$}
(in.east) edge[pil] (mu1.west);
\node[below=2.5cm of mu1] (mu4) {$\kv',\tilde\mu_4$}
(mu4.west) edge[pil] (in.east);
\node[right=2cm of pp] (mu3b) {};
\node[above=.1cm of mu3b] (mu3) {$-\kv,\tilde\mu_3$}
(in2.west) edge[pil] (mu3.east);
\node[below=2.5cm of mu3] (mu2) {$-\kv',\tilde\mu_2$}
(mu2.east) edge[pil] (in2.west);

 \node[right=6.5cm of market] (ladderd) {};
 \node[below=.75cm of ladderd] (ladder) {};
 \node[above=1cm of ladder] (l1) {};
 \node[right=3cm of l1] (l4) {}
 (l4.south) edge[electron] (l1.south);
 \node[below=2cm of l4] (l3) {};
 \node[below=2cm of l1] (l2) {}
 (l2.north) edge[electron] (l3.north)
 (l1.south) edge[photon] (l2.north)
 (l4.south) edge[photon] (l3.north);
  \node (pp) at ($(l1)!0.5!(l4)$) {};
  \node[below=.9mm of pp](){$\tilde \nu_2\quad~\pv+\kv+\kv'~\quad\tilde \nu_3$};
  \node[below=1.5cm of pp](){$\tilde \nu_1 ~\quad\qquad\pv ~\quad\qquad \tilde\nu_4$};
  
  \node[above=1.6cm of l1](l4in){$\kv',\tilde \mu_4$}
(l4in.south) edge[pil] (l4.south);
  \node[above=1.6cm of l4](l1out){$\kv,\tilde \mu_1$}
(l1.south) edge[pil] (l1out.south);

  \node[left=1cm of l2] (l2in) {};
  \node[above=.03cm of l2in] (l2inb) {$-\kv',\tilde \mu_2$}
(l2inb.south) edge[pil] (l2.north);
  \node[right=1cm of l3] (l3out) {};
  \node[above=.03cm of l3out] (l3outb) {$-\kv,\tilde \mu_3$}
(l3.north) edge[pil] (l3outb.south);

\end{tikzpicture}
\caption{Second order bubble and ladder diagrams. Note that each interaction line $U$ carries four joint indices $\tilde \mu=(\mu,s)$ for orbital and electronic spin.}
\label{fig:bubbleladder}
\end{figure*}

\section{Linearized gap equation and spin projection}

We denote the fermion operators $\beta_{n\spin}$ by band $n$ and pseudospin $\spin$. This is the natural basis when solving the linearized gap equation at the Fermi level. We construct 
 fermion bilinear operators 
 \begin{eqnarray}
 \overline\Psi_l(n,\kv)&=&s_l\beta^\dagger_{\kv n\spin_1}[\Gamma_l]_{\spin_1\spin_2}\beta^\dagger_{-\kv n'\spin_2}\delta_{n,n'} \nonumber \\
\Psi_l(n,\kv)&=&\beta_{\kv n\spin_1}[\Gamma_l]_{\spin_1\spin_2}\beta_{-\kv n'\spin_2} \delta_{n,n'},
\label{eq:Psi}
\end{eqnarray}
where $\spin$ denotes pseudo-spin, and the $[\Gamma_l]_{\spin_1\spin_2}$ matrices are constructed from the Pauli matrices $\sigma_l$ by
\begin{eqnarray}
 \Gamma_l&=&\frac{1}{\sqrt{2}}\sigma_l i \sigma_y .
\end{eqnarray}
Only intraband Cooper pairing is included, as implied by the $\delta$-function in Eqs.~(\ref{eq:Psi}).

Solutions are projected onto even-parity, pseudo-spin singlet $l=0$ and odd parity, pseudo-spin triplet with $l\in \{x,y,z\}$.
Here $s_0,s_y=-1$ and $s_x,s_y=+1$ and repeated indices are summed over. The $\Gamma$ spin matrices are thus given by
\begin{eqnarray}
 \Gamma_0&=&\frac{1}{\sqrt{2}}\sigma_0 i \sigma_y =\frac{1}{\sqrt{2}}\left[\begin{array}{cc}
                                                    0 & 1 \\
                                                    -1 & 0
                                                   \end{array}\right],\\
 \Gamma_x&=&\frac{1}{\sqrt{2}}\sigma_x i \sigma_y =\frac{1}{\sqrt{2}}\left[\begin{array}{cc}
                                                    -1 & 0 \\
                                                    0 & 1
                                                   \end{array}\right],\\
 \Gamma_y&=&\frac{1}{\sqrt{2}}\sigma_y i \sigma_y =\frac{1}{\sqrt{2}}\left[\begin{array}{cc}
                                                    i & 0 \\
                                                    0 & i
                                                   \end{array}\right],\\                                                   
 \Gamma_z&=&\frac{1}{\sqrt{2}}\sigma_z i \sigma_y =\frac{1}{\sqrt{2}}\left[\begin{array}{cc}
                                                    0 & 1 \\
                                                    1 & 0
                                                   \end{array}\right].                                               
\end{eqnarray}

To write the interaction Hamiltonian in terms of the projected operators, we  use the completeness relation;
\begin{eqnarray}
 \sum_{l=0}^3 s_l [\Gamma_l]_{\spin_1\spin_2}[\Gamma_l]_{\spin_3\spin_4}=\delta_{\spin_1\spin_4}\delta_{\spin_2\spin_3}
\end{eqnarray}
which is proven by use of
\begin{equation}
 tr \Gamma_i\Gamma_j=s_i\delta_{i,j} .
 \end{equation}
The interaction Hamiltonian is projected from orbital and electronic spin space, $(\mu,s)$, to band and pseudospin space $(n,\sigma)$. Thereafter the pairing vertex and fermion operators are projected to the pseudospin operators ($\Psi^l$) where $l=\{0,x,y,z\}$ refers to the definitions in Eq.~(\ref{eq:Psi}). In this manner the final interaction Hamiltonian takes the form:
\begin{widetext}

\begin{eqnarray}
 &&\hat H_{int} \nonumber \\
 &&= \frac{1}{2}\sum_{\kv,\kv',\{\mu\}}\Big[V(\kv,\kv')\Big]^{\mu_1 s_1 , \mu_2 s_2 }_{\mu_3 s_3,\mu_4 s_4} \quad c_{\kv \io_1 }^\dagger  c_{-\kv \io_3 }^\dagger c_{-\kv' \io_2 } c_{\kv' \io_4 }\nonumber \\
&&=\frac{1}{2}\sum_{\kv,\kv'\{n\}\{\spin\}} \underbrace{\sum_{\{\mu\}\{s\}} (u_{\mu_1 s_1}^{n_1 \spin_1}(\kv))^* (u_{\mu_3 s_3}^{n_1 \spin_3}(-\kv))^* 
\Big[V(\kv,\kv')\Big]^{\mu_1 s_1 , \mu_2 s_2 }_{\mu_3 s_3,\mu_4 s_4}  u_{\mu_2 s_2}^{n_2 \spin_2}(-\kv')  u_{\mu_4 s_4}^{n_2 \spin_4}(\kv')}_{[V(n_1,\kv;n_2,\kv')]^{\spin_1\spin_2}_{\spin_3\spin_4}} 
\quad \beta_{\kv n_1\spin_1}^\dagger \beta_{-\kv n_1\spin_3}^\dagger 
\beta_{-\kv' n_2\spin_2} \beta_{\kv' n_2\spin_4}\nonumber \\
&&\label{eq:cohf}\\
&=&\frac{1}{2}\sum_{ \kv,\kv' \{n\} \{\spin\}}
 \sum_{l,l'} 
 \underbrace{s_l \beta_{\kv n_1\spin_1}^\dagger [\Gamma_l]_{\spin_1\spin_3}\beta_{-\kv n_1\spin_3}^\dagger}_{\overline\Psi_{l}(n_1,\kv)}
 \quad\underbrace{\sum_{\{\tau\}} s_{l'}[\Gamma_l]_{\tau_3\tau_1}
[V(n_1,\kv;n_2,\kv')]^{\tau_1\tau_2}_{\tau_3\tau_4} 
[\Gamma_{l'}]_{\tau_4\tau_2}}_{\Gamma_{l,l'}(n_1 \kv;n_2 \kv')}
\quad\underbrace{\beta_{-\kv' n_2\spin_2} [\Gamma_{l'}]_{\spin_2\spin_4}
\beta_{\kv' n_2\spin_4}}_{\Psi_{l'}(n_2,\kv')} \nonumber \\
\end{eqnarray}
\end{widetext}
We ensure that the coherence factors $u_{\mu s}^{n \spin}(\kv)$ invoked in Eq. (\ref{eq:cohf}) when transforming the operators from orbital and spin space to band and pseudo-spin space do not carry random phases from the diagonalization process. Specifically, the eigenvectors aqcuired in the pseudo-spin up block diagonal must be related to the eigenvectors of the pseudo-spin down block diagonal by time-reversal symmetry due to Kramer's degeneracy. We ensure this by first numerically diagonalization one block-diagonal corresponding to pseudo-spin up and afterwards directly constructing the eigenvectors of pseudo-spin down by applying the time-reversal operator:
\begin{eqnarray}
\hat{T}=i\sigma_y \left(\begin{array} {ccc}
1 &0 &0 \\
0 &1& 0 \\
0 &0& -1 
\end{array}\right)\mathcal{K}
\end{eqnarray}
operating on $[d_{xz\up},d_{yz,\up},d_{xy,\down},d_{xz,\down},d_{yz,\down},d_{xy,\up}]$. The operator $\mathcal{K}$ implies complex conjugation. Furthermore, we impose the symmetry of $H(\kv)=H(-\kv)$ for the non-interacting Hamiltonian, which implies that  $u_{\mu s}^{n \spin}(\kv)=u_{\mu s}^{n \spin}(-\kv)$, to avoid a random gauge from each $\kv$ value.  By this procedure, the sum over pseudo-spins ensures that the possible random phases acquired through the numeric diagonalization are cancelled out.

The leading and sub-leading superconducting instabilities are determined from the common procedure by diagonalizing the matrix:
\begin{eqnarray}
  M_{\kv_f,\kv_f^\prime}=-\frac{1}{(2\pi)^2}\frac{l_{\kv_f^\prime}}{v(\kv_f^\prime)} \Gamma_{l,l'}(\kv_f,\kv_f^\prime),
  \label{eq:LGEs}
\end{eqnarray}
where $l_{\kv_f^\prime}$ is the length element of the Fermi surface and $v(\kv_f^\prime)$ is the Fermi velocity at $\kv_f^\prime$. Note that the pseudo-spin information of the vertex has been transferred to the indices $l,l'\in \{0,x,y,z\}$.
\begin{figure}[t]
 \centering
   	\includegraphics[angle=0,width=\linewidth]{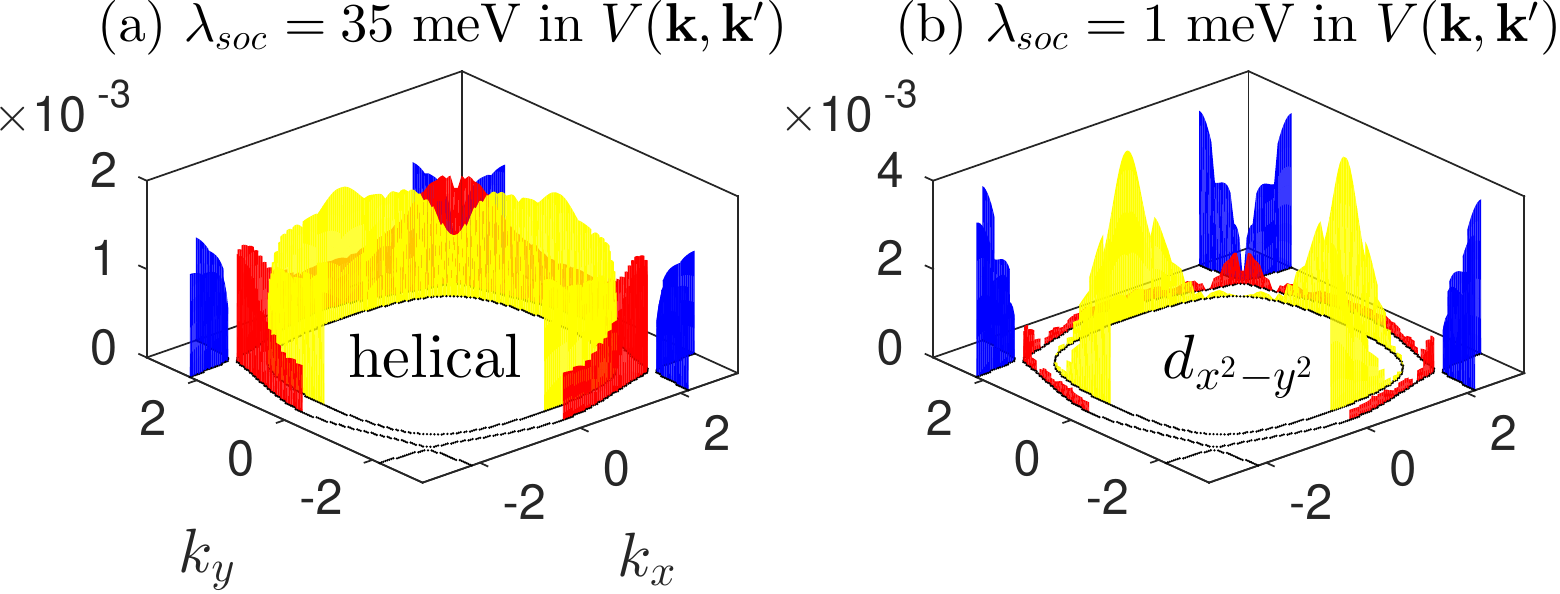}
\caption{Difference in leading gap solution when  (a) $\lambda_{soc}=35$ meV in the pairing kernel $V(\kv,\kv')$ of Eq.~(\ref{eq:cohf}) and when (b)  $\lambda_{soc}$ is decreased to $1$ meV exclusively in $V(\kv,\kv')$. Note that the helical solution appears only in the case of a sizable $\lambda_{soc}$ in the pairing kernel.}
\label{fig:Lam0}
 \end{figure}

\section{Effect of spin-orbit coupling in the pairing kernel}
In the phase diagrams shown in Fig. 1 of the main text, we observe odd-parity pseudo-spin triplet solutions only in the limit of large spin-orbit couplings, see Fig. 1 (b,d) of the main text. 
The appearance of these odd-parity states could occur either from the effect of spin-orbit coupling on the band structure, i.e. from the eigenvector elements $u_{\mu s}^{n \spin}(\kv)$ of Eq.~(\ref{eq:cohf}) or by more subtle effects from the presence of spin-orbit coupling in the spin susceptibilities entering the pairing kernel, i.e. $V(\kv,\kv')$ of Eq.~(\ref{eq:cohf}). A simple way to address this question is to keep a large spin-orbit coupling in the band construction, i.e. in $u_{\mu s}^{n \spin}(\kv)$, but diminish $\lambda_{soc}$ entering $V(\kv,\kv')$. In Fig.~\ref{fig:Lam0} we show the outcome of this approach; the helical solution results from the full calculation with $\lambda_{soc}=35$ meV, but when we set $\lambda_{soc}=1$ meV in the pairing kernel $V(\kv,\kv')$ a $d_{x^2-y^2}$ solution appears instead.
A similar conclusion holds for the chiral solution found at large $\lambda_{soc}$ in Fig. 1 (d) of the main text, where a $s^\prime$-wave solution appears instead of the chiral solution when $\lambda_{soc}=1$ meV in the pairing kernel. We conclude that the presence of SOC in the pairing kernel is crucial for the appearance of leading odd-parity solutions to the linearized gap equation.

\section{Density of States and Knight shift}

Calculation of the spectral gap is obtained by diagonalization of the BdG Hamiltonian at the Fermi surface:
\begin{eqnarray}
 \hat H_{MF}&=&\sum_\kv (\beta_{\kv\up}^\dagger \beta_{-\kv\down}\beta_{\kv\down}^\dagger \beta_{-\kv\up})H_\Delta(\kv)
 \left(\begin{array}{c}
 \beta_{\kv\up}  \\ \beta_{-\kv\down}^\dagger\\ \beta_{\kv\down} \\ \beta_{-\kv\up}^\dagger
 \end{array}\right), \nonumber \\
\end{eqnarray}
with
\begin{widetext}
\begin{eqnarray}
H_\Delta(\kv)&=&\left( \begin{array}{cccc}
 E(\kv)& \frac{1}{\sqrt{2}}(-\bar \Delta_0(\kv)+\bar \Delta_z(\kv))  & 0 &\frac{1}{\sqrt{2}}(-\bar \Delta_x(\kv)+i\bar \Delta_y(\kv)) \\
\frac{1}{\sqrt{2}}(-\Delta_0(\kv)+\Delta_z(\kv))  & -E(\kv)  & \frac{1}{\sqrt{2}}(\Delta_x(\kv)-i \Delta_y(\kv)) & 0 \\
  0 &\frac{1}{\sqrt{2}}(\bar \Delta_x(\kv)+i \bar\Delta_y(\kv))&E(\kv) & \frac{1}{\sqrt{2}}(\Delta_0(\kv)+\Delta_z(\kv)) 
  \\
  \frac{1}{\sqrt{2}}(-\Delta_x(\kv)-i \Delta_y(\kv))&0 & \frac{1}{\sqrt{2}}(\Delta_0(\kv)+\Delta_z(\kv)) & -E(\kv) 
 \end{array}\right), 
  \label{eq:BdG}
\end{eqnarray}
\end{widetext}
where $\kv$ is positioned at the Fermi surface and therefore $E(\kv)=0$. Also, we have therefore suppressed the band index, since it is uniquely defined by $\kv$. In the case of a pseudo-spin singlet or opposite pseudo-spin (chiral) triplet solution, the matrix Eq.~(\ref{eq:BdG}) becomes block-diagonal and the spectral gap is \begin{eqnarray}
\Delta_{\kv}=\frac{1}{\sqrt{2}}|\Delta_{(0/z)}(\kv)| .
\end{eqnarray}
In the case of two degenerate solutions $\Delta_x(\kv)$ and $\Delta_y(\kv)$, we solve the eigenvalue problem analytically in the case of $\Delta_{x/y}(\kv)$ purely real or purely imaginary. Thereby we obtain a spectral gap given by
\begin{eqnarray}
\Delta_{\kv}=\sqrt{\frac{1}{2}(|\Delta_x(\kv)|^2+|\Delta_y(\kv)|^2)}  .
\end{eqnarray}
\begin{figure}[t]
 \centering
   	\includegraphics[angle=0,width=.48\linewidth]{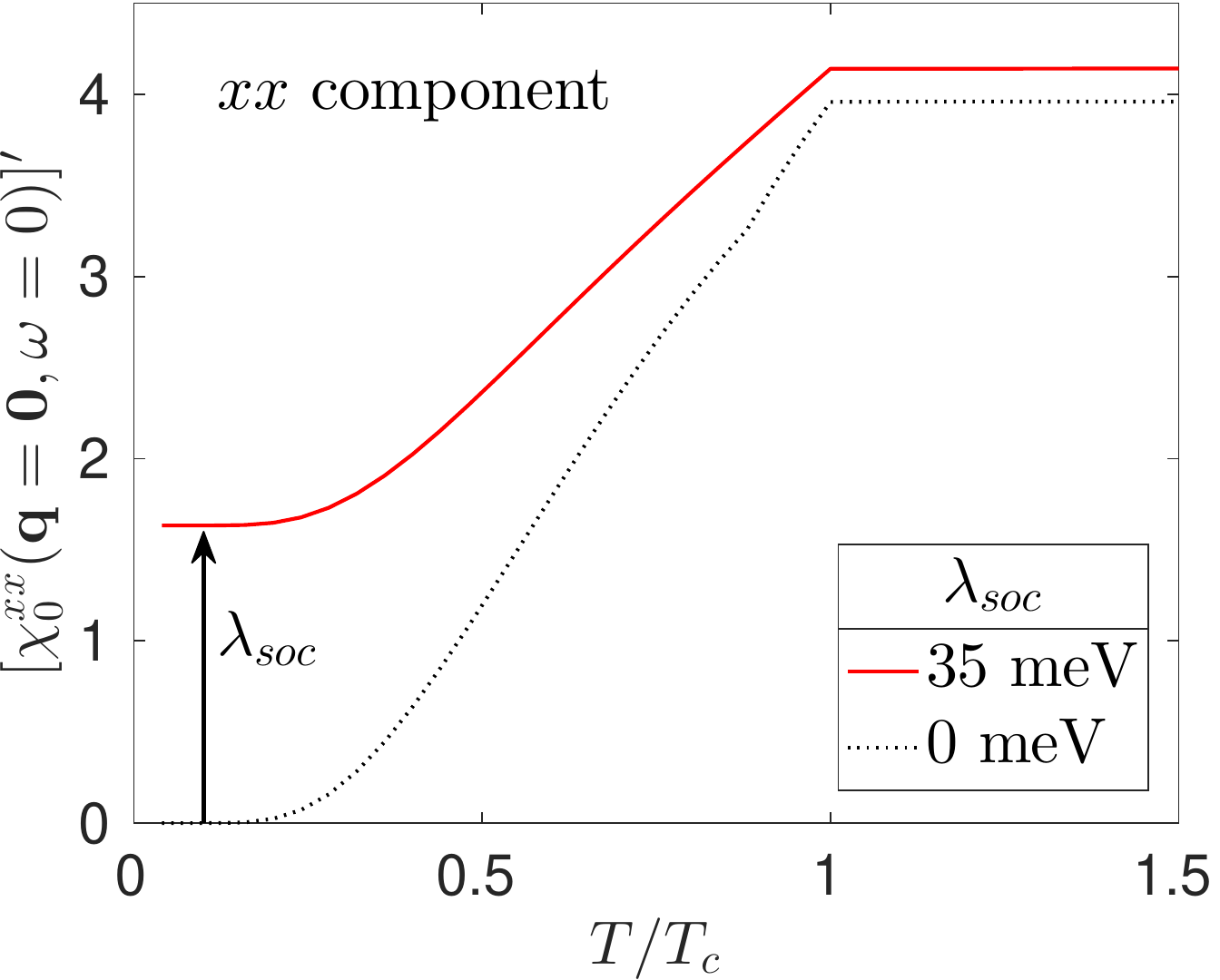}
   	\includegraphics[angle=0,width=.5\linewidth]{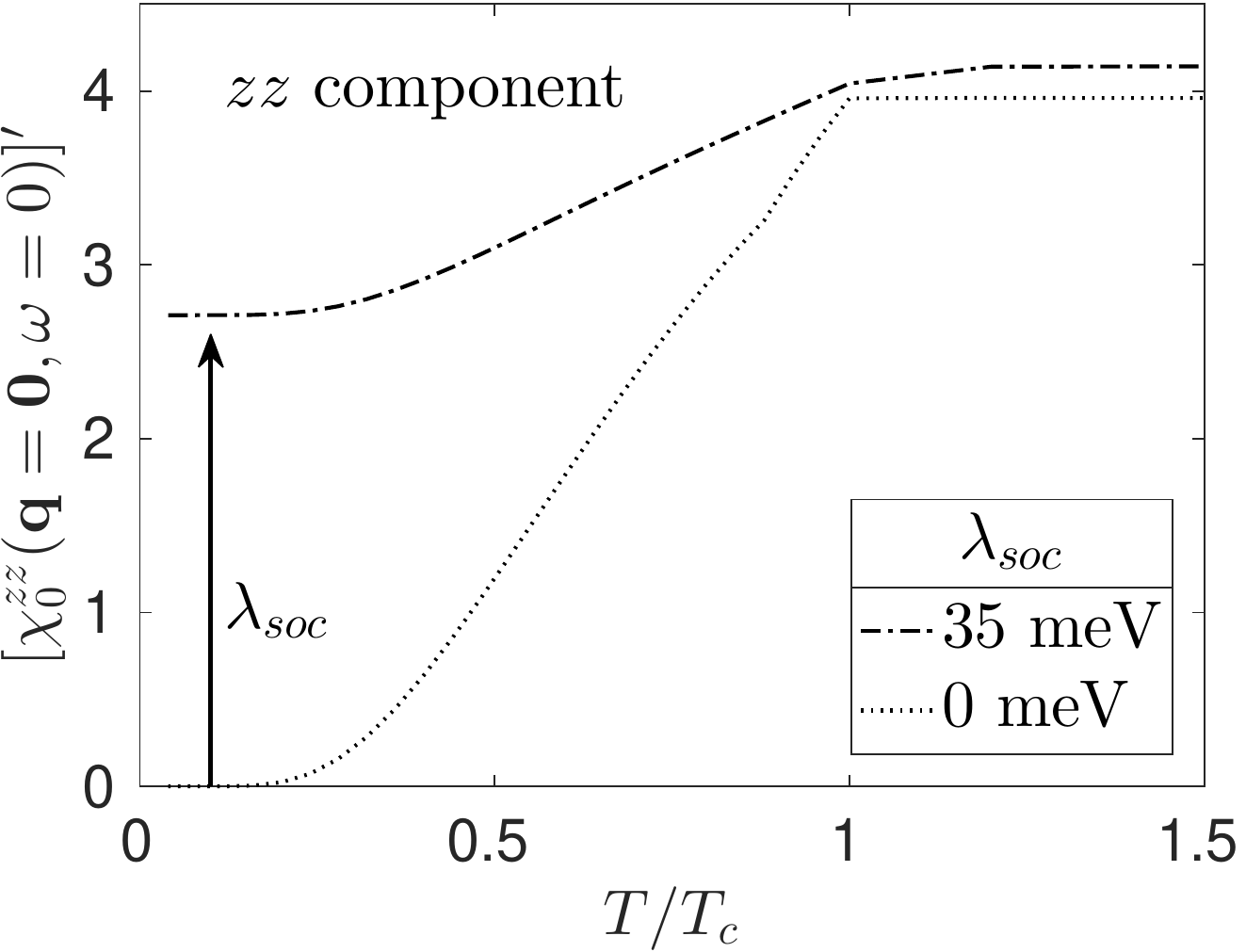}
\caption{Knight shift for a fully gapped conventional $s$-wave superconductor for (a) in-plane fields $xx$ and (b) out-of-plane fields $zz$ in the absence and presence of spin-orbit coupling, $\lambda_{soc}=35$ meV. The \sruo ~band is specified in the main text, setting $\mu=\mu_{xy}=109$ meV, and $k_B T_c=0.5$ meV with a uniform gap $\Delta(0)=1$ meV.}
\label{fig:knightswave}
 \end{figure}

The Hamiltonian ~(\ref{eq:BdG}) is the starting point for the calculation of the density of states and the spin susceptibility in the superconducting phase.
The matrix is diagonalized by a unitary transformation
 \begin{eqnarray}
 \beta_{n\kv\spin}            &=& a_{\kv n \spin}   \gamma_{n\kv\spin}  ~ + b_{\kv n \spin} \gamma_{n\kv\ospin}^\dagger    + c_{\kv n \spin}\gamma_{n\kv\ospin} + d_{\kv n \spin} \gamma_{n\kv\spin}^\dagger, \nonumber\\
 \beta_{n\kv\spin}^\dagger    &=& d^*_{\kv n \spin}   \gamma_{n\kv\spin}~ + c^*_{\kv n \spin} \gamma_{n\kv\ospin}^\dagger    + b^*_{\kv n \spin}\gamma_{n\kv\ospin} + a^*_{\kv n \spin} \gamma_{n\kv\spin}^\dagger,\nonumber\\
 \beta_{n-\kv\spin}^\dagger   &=& m_{\kv n \spin}   \gamma_{n\kv\spin}   + n_{\kv n \spin} \gamma_{n\kv\ospin}^\dagger    + o_{\kv n \spin}\gamma_{n\kv\ospin} + p_{\kv n \spin} \gamma_{n\kv\spin}^\dagger,\nonumber\\
 \beta_{n-\kv\spin}           &=& p^*_{\kv n \spin}   \gamma_{n\kv\spin}~ + o^*_{\kv n \spin} \gamma_{n\kv\ospin}^\dagger    + n^*_{\kv n \spin}\gamma_{n\kv\ospin} + m^*_{\kv n \spin} \gamma_{n\kv\spin}^\dagger. \nonumber\\
 \label{eq:BdGtrans}
 \end{eqnarray}
We calculate the spin- and orbital-resolved density of states $N_{\mu,s}(\omega)$ by
\begin{eqnarray}
 N_{\mu,s}(\omega)=-\sum_\kv Im G_{\mu, s}(\kv, \omega)
 \end{eqnarray}
where $G_{\mu, s}(\kv, \omega)$ is obtained from analytical continuation of
\begin{equation}
 G_{\mu, s}(\kv, i\omega_n)=-\int_0^\beta d \tau e^{i\omega_n\tau} \langle T_\tau c_{\kv\mu s}(\tau)c^\dagger_{\kv\mu s} \rangle. 
\end{equation}

By this we obtain by use of the transformations Eq.~(\ref{eq:cohf}) and Eqs.~(\ref{eq:BdGtrans})
\begin{eqnarray}
 &&N_{\mu,s}(\omega)=\nonumber \\
 &&- Im \sum_{\kv,n,\spin} |u^{n\spin}_{\mu s}(\kv)|^2 \Big[(|a_{\kv n \spin}|^2+|c_{\kv n \spin}|^2)\frac{1}{\omega-E_{\kv n}+i\eta}\nonumber\\
 &&\hspace{2.8cm}+(|b_{\kv n \spin}|^2 + |d_{\kv n \spin} |^2)\frac{1}{\omega+E_{\kv n}+i\eta}\Big]. \nonumber \\
 \end{eqnarray}
Here $a,b,c,d$ are eigenvector components and $E_{\kv n}$ is the eigenvalue of Hamiltonian (\ref{eq:BdG}).

The calculation of $\chi_0^{SC}$ departs from the expression

\begin{widetext}
\begin{eqnarray}
&& [\chi_0^{SC}]^{ \mu_1, s_1; \mu_2,s_2}_{\mu_3, s_3;\mu_4 s_4}(\qv,i\omega_n)= \frac{1}{N}\int_0^\gamma d \tau e^{i \omega_n \tau} \sum_{\kv} \sum_{n_1,n_2}\sum_{\spin_1,\spin_2,\spin_3,\spin_4}\nonumber  \\
&& 
\qquad U_{\mu_1 s_1}^{n_1 \spin_1*}(\kv-\qv) U_{\mu_2 s_2}^{n_2 \spin_2}(\kv) U_{\mu_3 s_3}^{n_2 \spin_3*}(\kv) U_{\mu_4 s_4}^{n_1 \spin_4}(\kv-\qv) 
\langle \beta^\dagger_{\kv-\qv,n_1,\spin_1} (\tau) \beta_{\kv-\qv,n_1,\spin_4} (0)\rangle_0 
\langle \beta_{\kv,n_2,\spin_2 }(\tau) \beta^\dagger_{\kv,n_2,\spin_3} (0) \rangle_0\nonumber  \\
&& 
\quad- U_{\mu_1 s_1}^{n_1 \spin_1*}(\kv-\qv) U_{\mu_2 s_2}^{n_2 \spin_2}(\kv) U_{\mu_3 s_3}^{n_1 \spin_3*}(-\kv+\qv) U_{\mu_4 s_4}^{n_2 \spin_4}(-\kv) 
\langle \beta^\dagger_{\kv-\qv,n_1,\spin_1} (\tau) \beta^\dagger_{-\kv+\qv,n_1,\spin_4} (0)\rangle_0 
\langle \beta_{\kv,n_2,\spin_2 }(\tau) \beta_{-\kv,n_2,\spin_3} (0) \rangle_0.\nonumber  \\
\label{eq:chiSC0}
 \end{eqnarray}
\end{widetext}
by use of the BdG transformation Eq.(\ref{eq:BdGtrans}).

In the linear response regime, we equate the Knight shift for the external magnetic field along $ \alpha \in\{x,y,z\}$  with the real part of the static spin-resolved susceptibility at $\qv=(0,0)$
\begin{eqnarray}
K^\alpha \propto Re [\chi^{\alpha \alpha}(\qv=0,i\omega_n=0)].
\end{eqnarray}
We leave out interactions and plot the susceptibility components $xx$, $yy$, and $zz$, labelled by electronic spins by use the expression for the physical susceptibility Eq.~(\ref{eq:chiphys}).
The bare susceptibility $[\chi_0]^{ \mu, s_1; \mu,s_2}_{\nu, s_3;\nu s_4}$ is calculated in the superconducting state by Eq.~(\ref{eq:chiSC0}). The temperature dependence of the gap magnitude is modelled by the BCS form:
\begin{equation}
\Delta(T)=\tanh\left[1.76\sqrt{\frac{T_c}{T}-1}\right]    
\end{equation}
The full superconducting gap is defined by:
\begin{eqnarray}
\Delta_l(\kv,T)=\Delta(T)\Delta_l (\kv)
\end{eqnarray}
where $\Delta_l(\kv)$ can be defined analytically or obtained from the leading eigenvector of the matrix defined in Eq.~(\ref{eq:LGE}), i.e. the solution to the linearized gap equation.
In the latter case, we extend the solution of the linearized gap equation to the full Brillouin zone in the following way: we assign for all $\kv$ vectors a gap $\Delta_l(\kv)$ which is given by the gap at the closest-lying Fermi wave vector damped by a Gaussian function.  
\begin{figure}[t]
\centering
   	\includegraphics[angle=0,width=\linewidth]{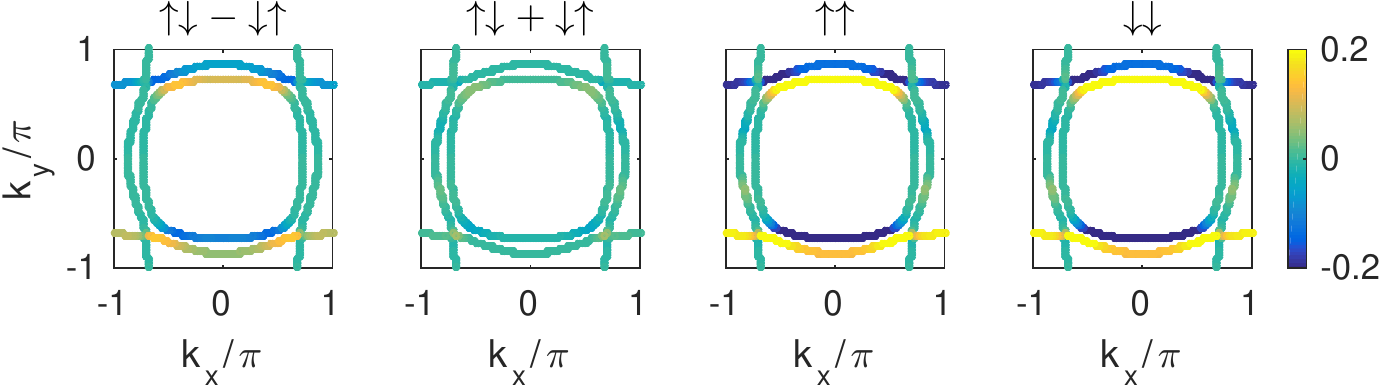}
   	\includegraphics[angle=0,width=\linewidth]{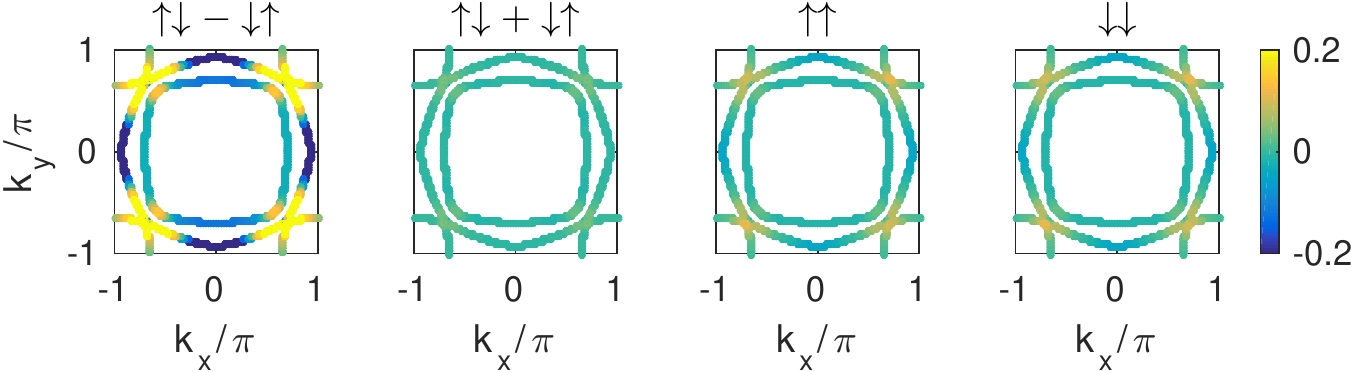}
   	\caption{Electronic structure of the leading gap solution of (upper row) band with $\mu=\mu_{xy}=109$ meV with $\lambda_{soc}=35$ meV, $U=120$ meV and $J/U=0.25$ (helical solution), (lower row) band with $\mu=90$ meV, $\mu_{xy}=128$ meV with $\lambda_{soc}=35$ meV, $U=100$ meV and $J/U=0.25$ (nodal $s^\prime$-wave). The helical solution displays strongest weight in the electronic same spin triplet channel, and a small weight in the electronic singlet channel while the nodal  $s^\prime$ solution shows strongest weight in the electronic singlet channel and small weight in the triplet channels. }
   	\label{fig:ElectronicCharacter}
\end{figure}
As a function of decreasing temperatures, the Knight shift shows signatures of spin-orbit coupling and the superconducting gap structure by differences between the three spin channels.
First, we consider the effect of a spin-orbit coupling of $\lambda_{soc}=35$ meV in a simple conventional $s$-wave gap. In the absence of spin-orbit coupling, the Knight shift is exponentially suppressed in all spin channels, see the dashed curves in Fig.~\ref{fig:knightswave}(a,b), as expected for a fully gapped singlet superconductor. In the presence of spin-orbit coupling, a residual Knight shift is present in all spin channels, and there is an additional difference between in-plane and out-of-plane spin directions. Especially for out-of-plane fields, the Knight shift suppression becomes much less pronounced, see Fig.~\ref{fig:knightswave} (b).
In general, when a system displays strong spin-orbit coupling the analysis of the Knight shift is complicated by the fact that an even-parity gap contains both singlet and triplet spin character, giving rise to residual Knight shifts at $T \to 0$ even for a fully gapped $s$-wave superconductor.

Another way to illustrate this is to transform the superconducting gaps back to electronic spin singlet and triplet character by 
\begin{widetext}
\begin{eqnarray}
 \langle c_{-\kv \mu_1 s_1}c_{\kv \mu_2 s_2}\rangle &=& \sum_{ \sigma_1 \sigma_2}u_{\mu_1 s_1}^{n_1 \sigma_1}(-\kv) u_{\mu_2 s_2}^{n_2 \sigma_2}(\kv) \langle \beta_{-\kv n \spin_1}\beta_{\kv n \spin_2}\rangle\nonumber \\
&=&  -u_{\mu_1 s_1}^{n \up}(-\kv) u_{\mu_2 s_2}^{n \up}(\kv) \frac{1}{\sqrt{2}} \langle \Delta_x(\kv) +i \Delta_y(\kv)\rangle +u_{\mu_1 s_1}^{n \down}(-\kv) u_{\mu_2 s_2}^{n \down}(\kv) \frac{1}{\sqrt{2}} \langle \Delta_x(\kv) -i \Delta_y(\kv)\rangle \nonumber \\
&&\quad+u_{\mu_1 s_1}^{n \up}(-\kv) u_{\mu_2 s_2}^{n\down}(\kv) \frac{1}{\sqrt{2}} \langle \Delta_0(\kv) + \Delta_z(\kv)\rangle +u_{\mu_1 s_1}^{n \down}(-\kv) u_{\mu_2 s_2}^{n \up}(\kv) \frac{1}{\sqrt{2}}\langle -\Delta_0(\kv) + \Delta_z(\kv)\rangle. 
\label{eq:Espinstruc}
\end{eqnarray}
\end{widetext}

This allows us to describe the gap solutions by labeling with real electronic spins. In Fig.~\ref{fig:ElectronicCharacter} the $\kv$-structure for the helical and $s^\prime$-wave gap in the electronic spin channels $(s_1,s_2)\in\{\up\down-\down\up,\up\down+\down\up,\up\up,\down\down\}$ is shown. We consider the Cooper pair $(-\kv \mu_1 s_1,\kv \mu_2 s_2)$ for $\kv$ at the Fermi surface ($\kv$ thus defines the band $n$ of the relevant transformation element $u_{\mu_1 s_1}^{n \sigma_1}(\kv)$ and the pseudospin is uniquely defined for a given pair $(\mu_1,s_1)$ ). For the results plotted in Fig.~\ref{fig:ElectronicCharacter}, we have summed in the orbital indices $(\mu_1,\mu_2)$ of Eq.~(\ref{eq:Espinstruc}).

\end{document}